\documentclass[11pt]{article}
\usepackage{latexsym}
\usepackage{graphicx}
	\graphicspath{{./}{figures}}
\usepackage{xcolor}
\usepackage{epstopdf}
\usepackage{amssymb}
\usepackage{amsmath,enumerate,rotate}
\usepackage{algorithm}
\usepackage{algpseudocode}
\usepackage{hyperref}

\def\Q{\hat Q}

\def\be#1\ee{\begin{equation}#1\end{equation}}

\newtheorem{remark}{\bf Remark}[section]

\newcommand{\bq}{\begin{equation}}
\newcommand{\eq}{\end{equation}}

\def\bqa{\begin{eqnarray}}
\def\eqa{\end{eqnarray}}

\def\l{\lambda}

\newcommand{\bd}{\begin{displaymath}}
\newcommand{\ed}{\end{displaymath}}
\newcommand{\ba}{\begin{eqnarray}}
\newcommand{\ea}{\end{eqnarray}}

\def\var{\varepsilon}

\newenvironment{equations}{\equation\aligned}{\endaligned\endequation}

\begin{document}

\title{Emergence of condensation patterns in kinetic equations for opinion dynamics}

\author{E. Calzola \thanks{Department of Computer Science, University of Verona, Strada Le Grazie 15, 37134 Verona, Italy. (elisa.calzola@univr.it)}  \and
	G. Dimarco\thanks{Department of Mathematics and Computer Science \& Center for Modeling, Computing and Statistics (CMCS), University of Ferrara, via Machiavelli 30, 44121 Ferrara, Italy. (giacomo.dimarco@unife.it)} \and
	G. Toscani\thanks{Institute of Applied Mathematics and Information Technologies, Via Ferrata 5, 27100 Pavia Italy.}\,\,\thanks{Department of Mathematics, University of Pavia, Via Ferrata 5, 27100 Pavia Italy. (giuseppe.toscani@unipv.it)}\and
	M. Zanella\thanks{Department of Mathematics, University of Pavia, Via Ferrata  5, 27100 Pavia Italy. (mattia.zanella@unipv.it)} 
}

\maketitle

\begin{abstract}
In this work, we define a class of models to understand the impact of population size on opinion formation dynamics, a phenomenon usually related to group conformity. To this end, we introduce a new kinetic model in which the interaction frequency is weighted by the kinetic density. In the quasi-invariant regime, this model reduces to a Kaniadakis-Quarati-type equation with nonlinear drift, originally introduced for the dynamics of bosons in a spatially homogeneous setting. From the obtained PDE for the evolution of the opinion density, we determine the regime of parameters for which a critical mass exists and triggers blow-up of the solution. Therefore, the model is capable of describing strong conformity phenomena in cases where the total density of individuals holding a given opinion exceeds a fixed critical size. In the final part, several numerical experiments demonstrate the features of the introduced class of models and the related consensus effects.
\end{abstract}


\tableofcontents

\section{Introduction}
Opinion formation is a complex process influenced by several factors including social interactions, personal experiences, cultural background, and exposure to information. Understanding how opinions form and evolve is crucial in many fields in social sciences \cite{opinion,turner91}. Mathematical models are able to provide powerful tools for studying opinion formation, allowing researchers in other disciplines to simulate and analyze different scenarios and to gain insights into the underlying mechanisms at the basis of the opinion consensus phenomena \cite{APTZ17,hegselmann2002opinion,PT13,To06}.

 In this work, we concentrate on the case where opinion formation dynamics are influenced by group conformity, that is intended as the influence in opinion formation dynamics exercised by an high number of agents sharing a certain similar opinion. One possible way to understand group conformity is as a form of social pressure. Indeed, understanding the human tendency to \emph{conform} to a larger group's opinion has become a point of interest of many disciplines, particularly in the field of social and psychological sciences. 
These concepts are considered as relevant also in political choices and voting behaviors \cite{Coleman_2004}. Already in 1958, Herbert Kelman \cite{kelman58} observed that there are different levels at which a change in attitude or actions of an individual can occur under the effect of social influence. These levels have different underlying processes leading the individual to accept the influence of the group because they want to cause a positive reaction from another person or group considered influential and/or avoid a punishment or disapproval. 
Social psychology works \cite{Asch1951,milgram63} mainly focused their studies on two type of conformity: normative conformity, that occurs when a person behaves so to be accepted by a larger group, and informational conformity, that happens when individuals adapt their opinion or behavior to the one of people believed to have some sort of authority or accurate information. In more recent years, scientists started to investigate the biological reasons behind this, and it is worth noticing that conformity is not only a social phenomenon, it may also has a \emph{biological} explanation. Neurosciences started investigating the brain's reaction to conformity suggesting an emotional response of the participants to the group's approval \cite{berns2005,stallen0}, even generating a reward response \cite{CampbellMeiklejohn2010HowTO}.

To shed light on this fascinating process and to try to reproduce qualitatively the above described conformism behaviors, in this work we consider a new model in which the size of a group  directly influences the dynamics of opinion formation. To this end, we will employ the methods of statistical physics, see e.g. \cite{Cer}. 
Recently, these techniques have proven to be highly effective and efficient tools for studying and analyzing opinion formation phenomena \cite{APTZ17,APZ14,AlettiNaldi,Dur09,FR,HRC,SPV,To06} and related aspects \cite{BTZ,braghini21,Burger1,CT18,DPTZ20,Dim21,Franceschi_Plos,KGFPS,Zan23}. { 
A key area of research focuses on modeling information exchange, with particular attention to understanding and analyzing how interpersonal influence impacts processes like opinion formation and the creation of connections. This aspect is closely linked to developing graph models for large networks \cite{burger22,during_24,fagioli24,nugent23}.} In this regard, one of the key concepts from statistical physics that can be applied to the dynamics of opinion formation is the notion of emergent behavior \cite{MT}, in which observable properties of large systems spontaneously  arise from interaction forces defined at the microscopic level. In this direction, kinetic Boltzmann-type equations are capable to naturally link microscopic agent-based interactions with their large time collective properties, elucidating the processes involved in consensus formation. 

In this work, we introduce a new microscopic mathematical description of opinion formation where each individual changes his own opinion as a consequence of the interactions with the whole population. The particular and innovative aspect of this interaction is that it is weighted with the intensity of the local density of agents sharing similar ideas. In our model, we also suppose that there is a certain amount of randomness in the interaction, modeling external factors which can be hardly controlled, such as the possibility to access information and the knowledge of every single individual. This microscopic dynamics is then inserted in a Boltzmann-type evolution equation and used to measure the variation in time of the density of individuals possessing an opinion on a given subject. Then, in order to try to get some insights on the behaviors of such dynamics we pass to the mean-field limit and we approximate the resulting Boltzmann equation through a grazing collision limit. This corresponds, from a modeling point of view, in assuming that the post-interaction opinions are very close to the pre-interaction ones, while at the same time the frequency of the interactions is suitably increased. The result obtained is a new Fokker-Planck-type model with nonlinear drift for the time evolution of the density of opinions, whose solution can present a blow-up to describe strong polarization phenomena due to the presence of group conformity. This blow-up appears only when the total density of individuals exceeds a fixed critical level and reveals strong analogies with some recent studies related to the Bose-Einstein condensate phenomenon \cite{ThiTo11,Pit03,Sp10,To11}. In this direction we mention the Fokker-Planck-type approach characterized by linear diffusion and superlinear drift introduced by Kaniadakis and Quarati  \cite{KQ93,Kania}{  for quantum indistinguishable particles}. 

In the second part of this work, we explore the new derived model from the numerical perspective point of view. Since both the microscopic as well as the Boltzmann dynamics are atypical, we first derive a new Monte Carlo method for the underlying  model \cite{Bao13,MC} and then we numerically show that indeed the Boltzmann equation and its Fokker-Planck approximation produce analogous solutions when the scaling is chosen in such a way that the grazing collision limit is approached by the Boltzmann dynamics. We successively use this new numerical method to document the capability of the model in describing strong polarization of opinions when a singular ensemble of individuals is considered and finally we extend the model to the case of distinct ensembles of individuals. This last case is constructed in such way to mimic political parties competition and the possible migration of people from one party to another during the political competition. For this last situation, we also derive conditions which permit to forecast the formation of strong consensus among one party during the time evolution of the underlying dynamics. 

The rest of the work is structured as follows. In Section \ref{sec:model} we present our new microscopic model of opinion formation and we derive the Boltzmann equation verified by the density of the opinions. In Section \ref{sec:Fokker} we approximate the Boltzmann type equation with a Fokker-Planck model passing to the grazing collision limit. We also compute the explicit analytical expression of the steady state and recover the value of the critical mass separating the case where the steady state remains smooth and the case where the steady state exhibits a blow-up. In Section \ref{sec:numerics}, we introduce a new Monte Carlo method and we use numerical simulations to validate the theoretical setting, showing in particular the resulting steady state density for various choices of the parameters and highlighting the differences in the shape of the distribution as the total mass increases or decreases, documenting blow-up situations. We also examine the case of several interacting populations of agents exchanging mass and opinions, as in the case of political factions interacting with different ideas on the same topic. We derive under which conditions one of the two populations reaches a critical mass and consequently a consensus emerges. In Section \ref{sec:conc}, we conclude the work with some future perspectives and open questions.

\section{Microscopic interactions for group conformity and consensus}\label{sec:model}
In this section, we propose a new model for opinion interactions within a population of homogeneous agents. This takes into account group conformity and leads to clusterization effects as discussed in the previous section. In recent years, a variety of heterogeneous social forces have been analyzed to understand, from a mathematical perspective, the process of consensus formation \cite{APTZ17}. In this direction, several models for opinions, based on kinetic theory \cite{Cer} have been introduced to describe the emergence of structures and patterns in real observed opinion dynamics, see e.g. \cite{DW15,To06} and the references therein. Those models have been often based on the assumption that the agents' opinion is modified by binary interactions with other individuals and possibly influenced by exogenous factors. In this direction, we mention recent works on the control of opinion formation dynamics \cite{APTZ17}, the case of interactions mediated by network topologies \cite{Calzola_2024,APZ17,LRT22,TTZ18} and multigroup dynamics, where agents may interact in a different fashion depending on the fact that they belong to a group or another, see e.g.  \cite{APZ14,Dur09}. Apart from the above recalled mechanisms, in a realistic context, there are more complex phenomena involved like stubbornness, competence, exposure to information, confirmation bias, heterogeneous influences or group dynamics, among others. The latter aspect is in particular taken into account in this work.

In the following, we comply with an alternative path and we are specifically interested by modeling how individuals modify their opinions in relation to the number of people sharing a certain common idea. This process can be thought as a consequence of psychological factors influencing decision making processes towards behavioral and group conformity patterns \cite{CY19,javarone}. To that aim, let us start with considering a population of indistinguishable agents characterized by their opinion $x \in \mathcal I = [-1,1]$, where the extreme values $-1$ and $+1$ represent, respectively, a strongly negative and a highly positive opinion on a specific subject. For simplicity we suppose that each agent modifies his opinion through interaction with a fixed background distribution $g(z):\mathcal I\to \mathbb R_+$ with 
\[
\int_{\mathcal I}z g(z)dz = m \in {(-1,1)},
\]
where $m$ is the mean opinion of the background and where this is supposed to mimic a distribution of opinion among a given population of individuals. We further introduce the distribution function $f = f(x,t):\mathcal I\times \mathbb R_+\to \mathbb R_+$ such that $f(x,t)dx$ represents the fraction of agents with opinion in $[x,x+dx]$ at time $t\ge0$. Hence, if $(x,z) \in \mathcal I\times \mathcal I$ is a pair of pre-interaction opinions, the post-interaction opinions $(x^\prime,z^\prime)$ is obtained 
by the following interaction scheme
\begin{equation}
	\label{eq:update}
	\begin{split}
		x^\prime &= x  + \epsilon P(x) (z-x) + D[f](x,t)\xi,\\
		z^\prime &= z,
	\end{split}
\end{equation}
where { $\epsilon \in (0,1/2]$}, $P(\cdot)\in [0,1]$ is an interaction function and $D[\cdot]\ge0$ represents the local relevance of the diffusion and it is such that $D[f](\pm 1,t) \equiv 0$ for all $t\ge0$. In addition, $\xi$ is a random variable with moments
\[
\left\langle \xi \right\rangle = 0, \qquad \textrm{and}\qquad
\left\langle \xi^2 \right\rangle = \frac{\epsilon}{\lambda}>0,
\]
{ with $\lambda > 0$}, having denoted with $\left\langle \cdot\right\rangle $ the expectation with respect to the random variable $\xi$.
The physical admissibility of the proposed interaction rule \eqref{eq:update} is guaranteed if $|x^\prime|\le 1$ for $|x|\le 1$, { meaning that, starting from an initial opinion $x \in \mathcal I$, the post-interaction opinion belongs to the $\mathcal I$. To this end, we can determine suitable bounds on the support of the random variable $\eta$, see also \cite{Cor05,TZ21,TZ_control}. We start by observing that} 
\[
\begin{split}
	|x^\prime|&\le (1-\epsilon P(x))x + \epsilon P(x)z + D[f](x,t)|\xi| \\
	&\le (1-\epsilon P(x))|x| +\epsilon P(x)  + D[f](x,t)|\xi|,
\end{split}\]
since $|x|\le 1$, from which we get that the sufficient condition to guarantee $|x^\prime|\le 1$ is provided by 
\[
D[f](x,t)|\xi|\le (1-\epsilon P(x))(1-|x|).
\]
This condition is satisfied if a constant $c>0$ exists and is such that 
\begin{equation}
	\label{eq:cond}
	\begin{cases}
		|\xi|\le c(1-\epsilon P(x)) \\
		cD[f] \le 1-|x|,
	\end{cases}
\end{equation}
for all $x,z\in \mathcal I$ and $f$. Hence, being $P\in [0,1]$ the first condition in \eqref{eq:cond} can be enforced by requiring 
\[
|\xi|\le c(1-\epsilon), 
\]
and it is sufficient to control the support of the random variable $\xi$ such that $|\xi|\le c(1-\epsilon)$. The second condition in \eqref{eq:cond} enforces $D[f](\pm 1,\cdot) = 0$ and it is needed to avoid opinions to exit from their support. In the linear case, where $D[f](x,t) = D(x)$, several choices for such local diffusion function have been analyzed in the past \cite{To06}. One possible form is $D(x) = 1-|x|$, which enforces the choice $c = 1$ and therefore the bound on the support of $|\xi|\le 1-\epsilon$. Another possibility is $D(x) = 1-x^2$, which enforces the choice $c = \frac{1}{2}$, and therefore $|\xi|\le \frac{1}{2}(1-\epsilon)$. 

Afterward, we will consider a different situation and we will deal with the following non linear case 
\begin{equation}
	\label{eq:D_choice}
	D[f](x,t) =\sqrt{ \dfrac{H(x)}{1+\beta (H(x) f(x,t))^\alpha}}
\end{equation}
where $H(\cdot)\ge0$ is a concave function such that $H(\pm 1) \equiv0$, and where the parameters $\alpha,\beta\ge0$ are introduced to weight the impact of the group conformity in the self-thinking process. In more details, { within the introduced choice, around the state $x \in \mathcal I$ self-thinking forces are mitigated by condensation effects which enforce compromise forces. In the following, we will concentrate on the following choices for the function $H(\cdot)$, weighting the local relevance of the diffusion: $H(x) = {1-x^2}$ and $H(x) = (1-x^2)^2$, see \cite{To06} for further details. In \eqref{eq:D_choice}, the parameter $\beta>0$ tunes the influence of nonlinear effects. In the following, we will determine the regime in which an opinion condensation occurs for sufficiently large values of $\alpha>0$.  }
{ Following now the analyses presented in \cite{PT13} we can observe that 
\[
\left\langle x^\prime + z^\prime \right\rangle = x + z + \epsilon P(x)(z-x),
\]
and consequently in such a model the mean opinion is not conserved on average unless $x\equiv z$ as expected since the distribution of $z$ is supposed not to vary with $x$. Furthermore, for $\epsilon \ll 1$ we have
\[
\left\langle (x^\prime)^2 + (z^\prime)^2 \right\rangle = x^2 + z^2 + 2\epsilon xP(x)(z-x) + \dfrac{D^2[f](x,t)}{\lambda} + o(\epsilon),
\]
and also the energy, playing the role of the variance with respect to the mean opinion, is not conserved on average in a single transition not even for $x\equiv z$. Looking now at the intensity of the transition we get
\begin{equation}
	\label{eq:variation_mean}
	\left\langle x^\prime - x \right\rangle = \epsilon P(x)(z-x)
\end{equation}
and 
\begin{equation}
	\label{eq:variation_second}
	\left\langle (x^\prime - x)^2 \right\rangle = \epsilon^2 P^2(x)(z-x)^2 + \epsilon \dfrac{D^2[f](x,t)}{\lambda  }.
\end{equation}}

The result of such process is that the compromise tendency becomes dominating in \eqref{eq:update}. The new rule for the change in opinions defined in \eqref{eq:update} can be now encoded in a Boltzmann-type kinetic equation which, in weak form, writes
\begin{equation}
	\label{eq:boltzmann}
	\begin{split}
	&\dfrac{d}{dt}\int_{\mathcal I}\varphi(x)f(x,t)dx =\\ &=\int_{\mathcal I \times \mathcal I}B_\epsilon[f](x,t) \left\langle \varphi(x^\prime)-\varphi(x)\right\rangle f(x,t)g(z)dx\,dz, 
\end{split}
\end{equation}
where $B_\epsilon[f](x,t)>0$ is a collision kernel. This equation describes the time evolution of the density of individuals sharing a given opinion where the ideas change in time with the law \eqref{eq:update}. The scope of the term $B_\epsilon[f](x,t)$ is to introduce a variable frequency enforcing larger interactions to highly populated opinions. The specific form chosen is the following 
\[
B_\epsilon[f](x,t) = 1+\beta(H(x)\min\left\{f(x,t),\epsilon^{-\alpha}\right\})^\alpha.
\]
This is introduced with the aim of modeling the fact that opinions with larger consensus find typically more space in discussions in particular on social platforms and media channels having consequently more possibilities to influence the population. In other words, the model rewards discussions where the leading general opinion is present, while comparatively disregards interactions in which the leading opinion is not present.

\begin{remark}
	One possible alternative way to model the effects of a strong exposure to leading information and social pressure in shaping the ideas of individuals could consist in modifying the interaction dynamics proposed in \eqref{eq:update} in the following way
	\begin{equation}\label{eq:micro2}
x' =\left\{
	\begin{array}{ll}
\min\{ x + (z-x)(1 + \beta (H(x)f)^\alpha) + H(x) \xi, 1\}, \quad & z>x
	 \\
\max \{ x + (z-x)(1 + \beta (H(x)f)^\alpha) + H(x) \xi, { - }1\}, \quad & z<x.
	\\ 
  	\end{array}
\right.
\end{equation}
	This different microscopic dynamics condenses the effects of all complex mechanisms involved in the formation of opinions through a mean field type of action where an agent align himself with the mean opinion expressed by the distribution of $g(z)$, i.e. the background fixed distribution of the society's idea. This alignment, as before, can be due to the fact that a person finds out about a topic over the net, reading newspapers, watching news, through social media connections or as a result of face-to-face discussions. To this effect, we add, as opposite to the previous interaction, a linear random variable which models the unpredictable role played by the environment and/or personal situation of the single individual such as beliefs, knowledge or influence. Finally, one weights the interactions with the size of the local density of people sharing a similar idea to characterize the propensity to conformity and the social pressure which leads individuals to align. Such type of microscopic dynamics leads to similar mesoscopic models as the one detailed in \eqref{eq:boltzmann}. 
\end{remark}

\section{Nonlinear Fokker-Planck kinetic models for group conformity}\label{sec:Fokker}
To get some insights about the model introduced in \eqref{eq:boltzmann} with microscopic opinion updates defined in  \eqref{eq:update}, here we develop an expansion techniques to reduce the complexity of the kinetic integro-differential equation \eqref{eq:boltzmann}. The idea consists in obtaining a Fokker-Planck-type partial differential equations in suitable regimes of parameters identified by the modeling choices which permits to mimic this slow gradual process of the formation of ideas and beliefs. This derivation has roots in the grazing collision limit \cite{FPTT12,Vill} and has been further explored in a variety of applications in socio-economic and life sciences in the recent past, see e.g. \cite{FPTT17,PT13}. 

Let $\varphi(x)$ be a test smooth function. From \eqref{eq:variation_mean}-\eqref{eq:variation_second} we get by Taylor expansion
\begin{equation}
	\begin{split}
	&\left\langle \varphi(x^\prime) - \varphi(x)\right \rangle = \dfrac{d}{dx}\varphi(x)\left \langle x^\prime -x\right\rangle \\
	&\qquad + \frac{1}{2} \dfrac{d^2}{dx^2}\varphi(x)\left\langle  (x^\prime-x)^2\right\rangle + \dfrac{1}{6}\dfrac{d^3}{dx^3}\varphi(\tilde x)\left\langle (x^\prime-x)^3\right\rangle,
\end{split}
\end{equation}
where $\tilde x \in (\min\{x^\prime,x\},\max\{x^\prime,x\})$. Plugging the above expansion in the model \eqref{eq:boltzmann} we get
\[
\begin{split}
	&\dfrac{d}{dt}\int_{\mathcal I}\varphi(x)f(x,t)dx = \\
	&\qquad\epsilon \int_{\mathcal I\times \mathcal I}B_\epsilon[f](x,t)\varphi^\prime(x)P(x)(z-x)f(x,t)g(z,t)dz\,dx + \\
	&\qquad\dfrac{\epsilon}{2\lambda}\int_{\mathcal I\times \mathcal I}B_\epsilon[f](x,t)\varphi^{\prime\prime}(x)D^2[f](x,t)f(x,t)g(z)dz\,dx +\\
	&\qquad R(f)(x,t),
\end{split}
\]
where $R(\cdot)$ is the so-called reminder term 
\begin{equation}
	\label{eq:reminder}
	\begin{split}
		&R(f)(x,t)  \\&= \dfrac{1}{2} \int_{\mathcal I\times \mathcal I}B_\epsilon[f](x,t)\dfrac{d^2}{dx^2}\varphi(x)\epsilon^2 P^2(x)(z-x)^2f(x,t)dx 
		\\&+ \dfrac{1}{6}\int_{\mathcal I\times \mathcal I}B_\epsilon[f](x,t)\left\langle (x^\prime-x)^3\right\rangle f(x,t)g(z)dx\,dz,
	\end{split}
\end{equation}
with
\[
\begin{split}
&\left\langle (x^\prime-x)^3 \right\rangle = \epsilon^3P^3(x)(z-x)^3\\ &\qquad+ \dfrac{3\epsilon^2}{\lambda} P(x)(z-x)D^2[f](x,t) + D^3[f](x,t)\left\langle \xi^3\right\rangle
\end{split}
\]
The modeling justification of the above expansion is now clear: we look for a regime where the single interactions between an individual and the background society are able only to slightly influence the personal believes. We look indeed to this process through a Taylor expansion and so to situations where $x'$ is close to $x$ as a result of a single exchange of ideas. Hence, by introducing the time scaling $\tau = t\epsilon$ and the scaled distribution $f_\epsilon(x,\tau) = f(x,t/\epsilon)$, we look now to the long time behavior of this process. This gives
\begin{equation}
	\label{eq:weak2}
	\begin{split}
		&\dfrac{d}{d\tau}\int_{\mathcal I}\varphi(x)f_\epsilon(x,\tau)dx  \\
		& =\int_{\mathcal I\times \mathcal I}B_\epsilon[f_\epsilon](x,\tau)\varphi^\prime(x)P(x)(z-x)f_\epsilon(x,\tau)g(z,t)dz\,dx  \\
		&+\dfrac{1}{2\lambda}\int_{\mathcal I\times \mathcal I}B_\epsilon[f_\epsilon](x,\tau)\varphi^{\prime\prime}(x)D^2[f_\epsilon](x,t)f_\epsilon(x,t)g(z)dz\,dx \\
		& +\dfrac{1}{\epsilon}R(f_\epsilon)(x,t).
	\end{split}
\end{equation}
Moreover, from \eqref{eq:reminder} we also get for $\epsilon\to 0^+$
\[
|R(f_\epsilon)| \lesssim (\epsilon^2 + \epsilon^3) \min\{f_\epsilon,\epsilon^{-\alpha}\}^\alpha  \to 0.
\]
Therefore, from \eqref{eq:weak2}, the limit $\epsilon \to 0^+$ gives
\[
\begin{split}
	&\dfrac{d}{d\tau}\int_{\mathcal I}\varphi(x)f_0(x,\tau)dx  \\
	&\qquad= \int_{\mathcal I\times \mathcal I}B_0[f_0](x,\tau)\varphi^\prime(x)P(x)(z-x)f_0(x,\tau)g(z)dz\,dx  \\
	&\qquad+\dfrac{1}{2\lambda}\int_{\mathcal I\times \mathcal I}B_0[f_0](x,\tau)\varphi^{\prime\prime}(x)D^2[f_0](x,t)f_0(x,t)g(z)dz\,dx.
\end{split}
\]
We can now integrating back by parts obtaining 
\[\begin{split}
&\partial_\tau f_0(x,\tau)\\ &= \partial_x \left[P(x)(x-m) f_0(x,\tau) B_0[f_0](x,t) +\dfrac{1}{2\lambda} \partial_x \left( H(x)f_0(x,\tau) \right)\right]
\end{split}
\]
thanks to an appropriate choice of the boundary conditions
\[
\begin{split}
	P(x)(x-m)f_0(x,\tau) B_0[f_0](x,t) +\dfrac{1}{2\lambda} \partial_x \left( H(x)f_0(x,\tau) \right) \Big|_{x = \pm 1} = 0 \\
	H(x)f_0(x,\tau) \Big|_{x = \pm 1} = 0
\end{split}
\]
affirming that there is no mass exiting from the boundaries of the domain, i.e. opinions can be at maximum extremely negative ($x=-1$) or extremely positive ($x=1$).

Coming back now for simplicity to the original variables, we observe that starting from \eqref{eq:boltzmann} we reduced ourselves to study the evolution of the distribution solution of a new nonlinear Fokker-Planck model reading 
\begin{equation}
	\label{eq:nonlinFP}
	\begin{split}
	&\partial_t f(x,t) = \partial_x \Big[ P(x)(x-m)(1+\beta(H(x)f(x,t))^\alpha)f(x,t)\\
	& +\dfrac{1}{2\lambda} \partial_x (H(x)f(x,t)) \Big].
\end{split}
\end{equation}
Interesting enough, the model obtained from this expansion shares many similarities with the well studied Kaniadakis-Quarati equation \cite{Kania,To11} with the main difference that in our case the solution lives in the bounded interval $\mathcal I$. The original Kaniadakis-Quarati model was indeed a Fokker-Planck equation with quadratic drift describing the dynamics of bosons exhibiting, in the spatially homogeneous setting, condensation behaviors. We may then expect that our model to be eventually able to generate in the long time similar behaviors.

We now notice that, under the assumption $P\equiv 1$ and $H(x) = 1-x^2$,  we get
\begin{equation}
	\label{eq:model1}\begin{split}
	&\partial_t f(x,t) = \partial_x \Big[(x-m)(1+\beta((1-x^2)f(x,t))^\alpha)f(x,t)+\\
	&+\dfrac{1}{2\lambda} \partial_x ((1-x^2)f(x,t)) \Big].
	\end{split}
\end{equation}
This Fokker-Planck model, in the absence of conformity forces, $\beta \equiv 0$, reduces to the usual Fokker-Planck model for opinion consensus in a bounded interval introduced in \cite{To06} and for which the large time distribution reads
\begin{equation}\label{smooth1}
	f_\infty(x) = c_{m,\lambda} \left(1+x \right)^{-1+\lambda(1+m)} \left(1-x \right)^{-1+\lambda(1-m)},
\end{equation}
with $c_{m,\lambda}>0$ a normalization constant. This equilibrium distribution corresponds to a beta-type distribution in the interval $\mathcal I$. Another possible choice, enforcing faster decay at the boundaries is $H(x) = (1-x^2)^2$, which, under the assumption $P\equiv 1$ gives
\begin{equation}
	\label{eq:model2}
\begin{split}
	\partial_t f(x,t) = &\partial_x\Big[ (x-m)(1+\beta((1-x^2)^2f(x,t))^\alpha)f(x,t)+ \\
	&+ \dfrac{1}{2\lambda} \partial_x ((1-x^2)^2 f(x,t)) \Big].
	\end{split}
\end{equation}
Therefore, in the absence of conformity forces, $\beta = 0$, \eqref{eq:model2} triggers the emergence of the equilibrium distribution
\begin{equation}\label{smooth2}
\begin{split}
	&f_\infty(x) = c_{m,\lambda} (1+x)^{-2+\lambda m/2}(1+x)^{-2-\lambda m/2}\exp\left\{-\dfrac{\lambda(1-mx)}{1-x^2}\right\},
\end{split}
\end{equation}
where, as before, $c_{m,\lambda}>0$ is a normalization constant. This second case depicts with respect to the case $H(x)=1-x^2$, a dynamics where strong positive or negative opinions can be handled with more difficulties by individuals in a population. 

In contrast to the smooth cases \eqref{smooth1} and \eqref{smooth2} characterizing dynamics of opinion formation in absence of mechanisms related to conformity, in this work, we are interested to the case of emergence of consensus when group pressures are active and may possibly modify the structure of the solution or in other words the ideas of individuals. For this reason, in the next section we will explore from the analytical point of view the specific cases $\alpha,\beta>0$ where the conformity forces are active leading to the new nonlinear Fokker-Planck model for opinion formation introduced in \eqref{eq:nonlinFP}. This study will highlight the different nature of our new approach introducing a new way to describe consensus phenomena not necessarily based on bounded confidence interactions \cite{hegselmann2002opinion}. 

\subsection{Blow up, existence of finite critical masses, steady states and entropy}
First of all, we want to prove the existence of a finite critical mass for the nonlinear Fokker-Planck model \eqref{eq:nonlinFP}. This quantity identifies the passage from a smooth steady state to a steady state presenting blow up. From a modeling point of view, the existence of a critical mass means that the size of the population plays a critical role in the consensus process. When, indeed the number of persons sharing a similar idea overtakes a threshold, consensus becomes unavoidable due to social pressure. We successively want to show that the steady state for this model corresponds to the unique minimizer of a suitable entropy functional. From the technical point of view, we follow in the sequel the approach proposed in \cite{abdallah2011minimization} in the case of condensates of bosons. 

Let observe now that equation \eqref{eq:nonlinFP} belongs to a more general class of Fokker-Planck equations having the form 
\begin{equation}
	\label{eq:nonlinFP_general}
	\partial_t f(x,t) = \partial_x \left[ (x-m) \mathcal D[f] + \sigma^2 \partial_x \mathcal R[f]\right], 
\end{equation}
being $\mathcal D[f]>0$ the  drift functional and $\mathcal R[f]$ a diffusion functional. This general form will be then used in the following to characterize the solution. We start now by rewriting equation \eqref{eq:nonlinFP} for $x \in (-1,1)$ with respect to $g(x,t) =  H(x)f(x,t)$ to get
\begin{equation}\label{eq:fpbenew}
	\begin{split}
	&\dfrac{1}{H(x)}\partial_tg(x,t) \\
	&\quad= \frac{\partial}{\partial x}\left[\frac{x-m}{H(x)}g(x,t)(1 + \beta g^\alpha(x,t)) +\dfrac{1}{2\lambda} \partial_x g(x,t)\right], 
\end{split}
\end{equation}
from which we get that the unique steady state $g_\infty(x)$, if it exists, is such that 
\[
2\lambda\frac{x-m}{H(x)}g_\infty(x)(1 + \beta g_\infty^\alpha(x)) + \partial_x g_\infty(x)=0, 
\]
which gives
\begin{equation}
	\label{eq:equa_ginfty}
	\partial_x \log\dfrac{g_\infty(x)}{(1+\beta g_\infty^\alpha(x))^{1/\alpha}} = -2\lambda\dfrac{x-m}{H(x)}.
\end{equation}
It is therefore of interest to understand the condition under which $g^\infty(x)$ solution to \eqref{eq:equa_ginfty} is a  density function, i.e. $g^\infty(x)\geq 0$, additionally weighted by a given mass $M$ defined later. To this end, we observe that \eqref{eq:nonlinFP} recast in the form \eqref{eq:nonlinFP_general} gives as drift functional
\begin{equation*}
	\mathcal D[g] = g(1 + \beta g^\alpha),
\end{equation*}
where $\mathcal D[g]>0$ for all $g >0$. We observe also that $\mathcal D[g]$ is superlinear, meaning that
\begin{equation}
	\label{eq:cond_drift}
	\int_1^\infty \frac{1}{\mathcal D(\rho)}{d}\rho \leq K <\infty.
\end{equation}
The steady state of \eqref{eq:fpbenew} of given mass $\mu=\int_{\mathcal I}f(x,t)dx>0$ solves then
\[
\dfrac{1}{2\lambda}\partial_x g + \dfrac{x-m}{H(x)}\mathcal D[g] =0,
\]
which can be rewritten as
\begin{equation}
	\label{condition}
	\dfrac{1}{\mathcal D[g]}\partial_x g +2\lambda \partial_x V(x) =0,
\end{equation}
supposing that a $V(x)\ge0$ such that 
\[
\dfrac{d}{dx}V(x) = \dfrac{x-m}{H(x)},
\]
exists. To continue the computation we need now to specify the shape of the function $V(x)$. To that aim, the two leading examples for the shape of the concave function $H(x)$, which scope is to make the diffusion disappear close to the boundaries, are analyzed. In the case $H(x)= 1-x^2$, we get
\[
V(x) = \log \left[ (1-x)^{\frac{m-1}{2}}(1+x)^{-\frac{m+1}{2}}\right], 
\]
whereas for $H(x) = (1-x^2)^2$ we get
\[
V(x) = \dfrac{1}{4} \left( \dfrac{2(mx-1)}{x^2-1} + \log\left( \dfrac{1-x}{1+x}\right)^m\right).
\]
We define now $\Phi'(g)$ as
\begin{equation}\label{eq:phip}
	\Phi^\prime(g) = -\int_g^\infty \frac{1}{\mathcal{D}(\rho)}d\rho, 
\end{equation}
which gives, in the case $\mathcal D[g] = g(1+\beta g^\alpha)$
\begin{equation}
	\label{eq:phiprime1}
	\Phi^\prime(g) =\dfrac{1}{\alpha} \log \dfrac{g^\alpha}{1+\beta g^\alpha}. 
\end{equation}
Substituting now $\Phi(\cdot)$ into \eqref{condition}, we obtain that the steady state is solution to 
\begin{equation}\label{eq:phi}
	\partial_x \left[ \Phi^\prime(g) +2\lambda V(x) \right] = 0, 
\end{equation}
or equivalently
\begin{equation}
	\label{eq:phiprime2}
	\Phi^\prime(g) = -2\lambda V(x)-c, 
\end{equation}
with $c$ a constant such that $c>0$. Since $\Phi^\prime(\cdot)$ is strictly increasing, denoting by $[\Phi^\prime]^{-1}$ its inverse, we can alternatively express the steady state of \eqref{eq:fpbenew} as follows
\begin{equation}
	\label{eq:st2}
	g_{\infty,c} = [\Phi^\prime]^{-1}\left( -2\lambda V(x)-c  \right). 
\end{equation}
As a consequence of \eqref{eq:phip}, we have that
\[
\Phi(g) = \int_0^g \Phi^\prime(\rho)d\rho,
\]
where we observe that the function $\Phi(\rho)$, which has a derivative that is non-decreasing, is then convex for $\rho\ge0$. This latter observation permits therefore to consider an entropy functional associated to the steady state solution\eqref{eq:phi}
\[
\mathcal H[g](t) = \int_{\mathcal I}\left( 2\lambda V(x)g + \Phi(g) \right)\dfrac{1}{H(x)}dx,
\]
which has the property to decrease in time along the solution to equation \eqref{eq:fpbenew}. In fact we have that 
\[
\begin{split}
	\frac{d}{dt} \mathcal{H}[g] (t) =& \int_{-1}^1 \left( 2\lambda V(x) + \Phi'(g)\right)\dfrac{1}{H(x)}\partial_t g \;dx \\
	=& \int_{-1}^1 \left( 2\lambda V(x) + \Phi'(g)\right)\frac{\partial}{\partial x}\left\{\mathcal D[g] \frac{\partial}{\partial x}\left[2\lambda V(x) + \Phi'(g)\right]\right\}\;dx \\
	=&-\int_{-1}^1 \mathcal D[g]\left( \frac{\partial}{\partial x}\left[2\lambda V(x) + \Phi'(g)\right]\right)^2\;dx \leq 0.
\end{split}
\]
{ Thus, we have obtained that, starting from an initial density with a given mass $\mu$, the solution is such that $\mathcal H[\cdot]$ is dissipated monotonically. Formally, we have obtained that the steady state is the one given by \eqref{eq:st2} which, by construction, preserves the initial mass $\mu$ over time.} 

We can distinguish now two different situations depending on the initial value of $\mu$. The first case refers to a smooth equilibrium distribution, while the second case refers to a solution presenting a blow-up where concentration of a part of the total mass around the mean opinion $m=\int_{\mathcal I}xf(x,t)dx$ can be observed. To that aim, we focus now on the specific case $m=0$. The steady state solution reads in this situation 
\[
g_{\infty,c}(x) = \dfrac{e^{-2\lambda V(x)-c}}{\left(1-\beta e^{-2\lambda V(x)-c}\right)^{1/\alpha}},
\] 
which may eventually have a critical state and thus presents a blow-up, only if $1-\beta e^{-c}e^{-2\lambda V(x)}=0$. In the case $H(x)=1-x^2$, this implies that the constant in equation \eqref{eq:phiprime2} has to take the  value $c = \log\beta$, being also $x = 0$ the unique minimum of $V(x)\ge0$ with $V(0) = 0$. On the other hand, in the case $H(x) = (1-x^2)^2$ and $m=0$, with similar computations one gets $c = \log \beta - \lambda$.  We want now to find the values of the parameters in \eqref{eq:st2} such that the integral of $g_{\infty,c}$ is finite, in other words the solution found has a physical sense.
To that aim, let us go back to \eqref{eq:st2} with $H(x) = 1-x^2$ and $m=0$. In this situation, we have that the critical mass $\mu_c(g_{\infty})$ is
\[
\begin{split}
	\mu(g_{\infty}) &= \int_{-1}^1 g_{\infty,\log\beta} \;{d}x \\
	&= \int_{-1}^1 \left[ \Phi'\right]^{-1} \left(\lambda\log \left( 1-x^2\right) -\log\beta \right)dx \\
	&= 2\int_{0}^1 \left[ \Phi'\right]^{-1} \left(\lambda\log ( 1-x^2)-\log\beta  \right)dx.
\end{split}\]
Following \cite{abdallah2011minimization} we change the variables in the above integral with $s$ so that $\Phi'(s) = \lambda\log(1-x^2)-\log\beta$ and we get
\begin{equation}\label{eq:mc}
	\begin{split}
		\mu(g_{\infty}) =&-\dfrac{1}{\lambda}\int_{+\infty}^{0}\dfrac{ s \left(\beta e^{\Phi^\prime(s)}\right)^{1/\lambda}\Phi^{\prime\prime}(s)}{\sqrt{1-\left(\beta e^{\Phi^\prime(s)}\right)^{1/\lambda}}}ds \\
		=& \dfrac{1}{\lambda} \int_0^{+\infty}\dfrac{ 
		\beta^{1/\lambda} \left(\dfrac{s^\alpha}{1+\beta s^\alpha}\right)^{\frac{1}{\lambda\alpha}}}{(1+\beta s^\alpha)\sqrt{1-\beta^{1/\lambda} \left(\dfrac{s^\alpha}{1+\beta s^\alpha}\right)^{\frac{1}{\lambda\alpha}}}} ds.
	\end{split}
\end{equation}
The integrand in the last integral appearing in \eqref{eq:mc} as $s\to \infty$ behaves like $s^{-\alpha/2}$, so there exists a finite critical mass only if $\alpha > 2$.

A similar computation can be done in the case $H(x) = (1-x^2)^2$ and $m = 0$. In this case, the critical mass $\mu(g_{\infty})$ related to the drift $\mathcal D(g) = g(1+\beta g^\alpha)$ is obtained as
\[
\begin{split}
	\mu(g_{\infty}) &= \int_{-1}^1 g_{\infty,\log\beta-\lambda}dx = \int_{-1}^1 [ \Phi^\prime]^{-1}\left(-\dfrac{\lambda}{x^2-1}-\log\beta + \lambda\right)dx \\
	& = 2\int_0^1  [ \Phi^\prime]^{-1}\left(-\dfrac{\lambda}{x^2-1}-\log\beta + \lambda\right)dx.
\end{split}
\]
Hence, we perform the change of variables from $\Phi^\prime(s) = -\dfrac{\lambda}{x^2-1}-\log\beta+\lambda$ and we get
\begin{equation}
	\label{eq:mc2}
	\begin{split}
		\mu(g_{\infty}) &=  \int_{0}^{+\infty} \dfrac{\lambda\Phi^{\prime\prime}(s)s}{\left(\Phi^\prime(s) + \log\beta - \lambda \right)^{3/2}} \dfrac{1}{\sqrt{\Phi^\prime(s) + \log\beta }}ds \\
		&= \int_0^{+\infty} \dfrac{\lambda  \left( \log \dfrac{\beta s}{(1+\beta s^\alpha)^{1/\alpha}}  \right)^{-1/2}}{(1+\beta s^\alpha)\left( \log \left( \dfrac{s^\alpha}{1+\beta s^\alpha}\right)^{1/\alpha} + \log\beta - \lambda\right)^{3/2}} ds,
\end{split}\end{equation}
where the integrand behaves as $s^{-\alpha/2}$ for $s \to +\infty$ and a critical mass exists, as before, only if $\alpha>2$. 
To conclude, as stated in the introduction, our new model in the case $m=0$ and for values of the function $H(x)=1-x^2$ or $H(x)=(1-x^2)^2$ presents blow-up when the total density of individuals exceeds a fixed critical level and $\alpha>2$. This result will be used in the numerical section to describe strong polarization phenomena due to the presence of social pressure.

\subsection{Critical value for non centered steady states}
In the case $m\ne 0$, the computations done in the previous section which lead to the existence of a finite critical mass cannot be explicitly done. Thus, in what follows we make use of a different argument to prove an analogous result. In particular we start from the explicit expression of the steady state obtained from \eqref{eq:fpbenew}.

Let us first consider the case $H(x) = 1-x^2$ . To this extent, consider that \eqref{condition} implies that the distribution $g$ of the stationary state satisfies the differential equation
\begin{equation}
	\frac{1}{g(1+\beta g^\alpha)}\partial_x g = -2\l\frac{x-m}{1-x^2}.
\end{equation}
Since
\[
\frac{1}{g(1+\beta g^\alpha)} = \frac{1}{g} - \frac{1}{\alpha} \frac{\beta \alpha g^{\alpha-1}}{1+\beta g^\alpha},
\]
and
\[
\frac{1}{g(1+\beta g^\alpha)}\partial_x g= \partial_x  \log\left( \frac{g}{(1+\beta g^\alpha)^{1/\alpha}}\right),
\]
we get
\[
\partial_x  \log\left( \frac{g^\alpha}{1+\beta g^\alpha}\right) = -2\alpha\lambda \partial_x \log\left[ (1+x)^{-\frac{m+1}{2}}(1-x)^{\frac{m-1}{2}}\right].
\]
Therefore
\begin{equation}\label{eq:ginf}
	\frac{g^\alpha}{(1+\beta g^\alpha)} = C^\alpha(1+x)^{\l\alpha(1+m)}(1-x)^{\l\alpha(1-m)},
\end{equation}
that implies
\[
g_{\infty,C}= \frac{C(1+x)^{\l(1+m)}(1-x)^{\l(1-m)}}{(1-\beta C^\alpha(1+x)^{\l\alpha(1+m)}(1-x)^{\l\alpha(1-m)})^{1/\alpha}},
\]
and, finally
\begin{equation}
\label{finf}
f_{\infty,C} = \frac{C(1+x)^{\l(1+m)-1}(1-x)^{\l(1-m)-1}}{(1-\beta C^\alpha(1+x)^{\l\alpha(1+m)}(1-x)^{\l\alpha(1-m)})^{1/\alpha}}.
\end{equation}
Now, it is important to remark that the recovered steady state must be a non negative function. This means that, for  a given  positive constant $C$ the denominator of the fraction in \eqref{finf} must be non negative. This implies an upper bound on the possible values taken by the constant $C$. In this regard it is sufficient to note that for $g \ge 0$,  $g^\alpha/(1+\beta g^\alpha) \leq 1/\beta$.  Consequently the constant $C$ appearing in \eqref{eq:ginf} has to satisfy the condition
\[
C(1+x)^{\l(1+m)}(1-x)^{\l(1-m)} \leq \left(\frac{1}{\beta}\right)^{{1}/{\alpha}}.
\]
On the other hand,  the function $(1+x)^r(1-x)^s$, for $\left|x\right| \leq 1$, takes the maximum in $\overline  x = (r-s)/(r+s)$, so that
\[
(1+x)^r(1-x)^s \leq \left( \frac{2r}{r+s}\right)^r \left( \frac{2s}{r+s}\right)^s.
\]
In our case, since $r=\l(1+m)$, $s=\l(1-m)$, the maximum value is attained at $\overline x = m$ which implies the condition
\[
C(1+m)^{\l(1+m)}(1-m)^{\l(1-m)} \leq \left(\frac{1}{\beta}\right)^{\frac{1}{\alpha}}.
\]
Finally, the constant $C$ in \eqref{finf} can not be greater than $C_M$, given by
\[
C_M = \left(\frac{1}{\beta}\right)^{\frac{1}{\alpha}}(1+m)^{-\l(1+m)}(1-m)^{-\l(1-m)}.
\]
In correspondence to this value, the steady state $f_{\infty,{C_M}}(x)$ is such that 
\[
\lim_{x\to m} f_{\infty,{C_M}}(x) = + \infty,
\]
i.e. we observe a blow-up. Moreover, a direct computation shows that 
\[
\lim_{x\to m} \frac{ 1-\beta C_M^\alpha(1+x)^{\l\alpha(1+m)}(1-x)^{\l\alpha(1-m)}}{(x-m)^2} = \lambda\alpha(1-m^2)^{-1},
\]
which implies that the steady state $f_{\infty,{C_M}}$ tends to infinity at the order $2/\alpha$ as $x \to m$ as for the case $m=0$. Consequently, 
the integral  
\begin{equation}\label{eq:massacritica}
	\mu_c = \int_{-1}^{1} f_{\infty,{C_M}}(x)\mathrm{d}x
\end{equation}
is bounded as soon as $\alpha >2$. Therefore, as $\alpha >2$  we conclude with the existence of a finite critical mass $\mu_c$, coherently with what proven in \eqref{eq:mc}. 
Next, we consider the case $H(x) = (1-x^2)^2$. From \eqref{condition} we get that the distribution $g$ of the stationary state is such that 
\[
\dfrac{\partial_x g}{g(1+\beta g^\alpha)} = -2\lambda \dfrac{x-m}{(1-x^2)^2}. 
\]
Hence, proceeding as before, we have
\[
\partial_x \log\left(\dfrac{g^\alpha}{1+\beta g^\alpha} \right) = -\alpha\lambda \partial_x \left[ \dfrac{1-mx}{1-x^2} +\dfrac{1}{2}\log \left( \dfrac{1-x}{1+x}\right)^m\right],
\]
and therefore
\[
\dfrac{g^\alpha}{1+\beta g^\alpha} = C^\alpha e^{-\alpha\lambda \frac{1-mx}{1-x^2}}\left(\dfrac{1-x}{1+x}\right)^{-\frac{\alpha\lambda m}{2}},
\]
which implies 
\begin{equation}
	\label{eq:ginf2}
	g_{\infty,C} = \dfrac{Ce^{-\lambda \frac{1-mx}{1-x^2}}\left(\frac{1-x}{1+x}\right)^{-\frac{\lambda m}{2}}}{1-\beta C^\alpha e^{-\alpha\lambda \frac{1-mx}{1-x^2}}\left( \frac{1-x}{1+x}\right)^{-\frac{\alpha\lambda m}{2}}},
\end{equation}
and, finally
\begin{equation}
	\label{eq:finfty2}
	f_{\infty,C} = (1-x^2)^{-2} \dfrac{Ce^{-\lambda \frac{1-mx}{1-x^2}}\left(\frac{1-x}{1+x}\right)^{-\frac{\lambda m}{2}}}{\left(1-\beta C^\alpha e^{-\alpha\lambda \frac{1-mx}{1-x^2}}\left( \frac{1-x}{1+x}\right)^{-\frac{\alpha\lambda m}{2}}\right)^{1/\alpha}}. 
\end{equation}
To determine the critical value of the constant $C$ in \eqref{eq:ginf2} we get as before the condition
\[
C e^{-\lambda \frac{1-mx}{1-x^2}}\left(\dfrac{1-x}{1+x}\right)^{-\frac{\lambda m}{2}} \le \left(\frac{1}{\beta} \right)^{1/\alpha},
\]
whose maximum is attained at $x = m$, which implies the condition
\[
C e^{-\lambda }\left(\dfrac{1-m}{1+m}\right)^{-\frac{\lambda m}{2}} \le \left(\frac{1}{\beta} \right)^{1/\alpha}.
\]
Therefore, the constant $C>0$ cannot be greater than $C_M$ given by 
\[
C_M = \left(\frac{1}{\beta} \right)^{1/\alpha}e^{\lambda }\left(\dfrac{1-m}{1+m}\right)^{\frac{\lambda m}{2}}, 
\]
for which the steady state $f_{\infty,C_M}(x)$ is such that 
\[
\lim_{x \to m}f_{\infty,C_M}(x) = +\infty. 
\]
Furthermore, we get
\[
\lim_{x\to m} \dfrac{1-\beta C_M^\alpha e^{-\alpha\lambda \frac{1-mx}{1-x^2}}\left( \frac{1-x}{1+x}\right)^{-\frac{\alpha\lambda m}{2}}}{(x-m)^2} = \alpha \lambda (1-m^2)^{-2}, 
\]
implying that also the steady state $f_{\infty,C_M}$ defined in \eqref{eq:finfty2} tends to infinity at the order $2/\alpha$ as soon as $x \to m$ and the integral 
\[
\mu_c = \int_{-1}^1 f_{\infty,C_M}dx
\]
is bounded for $\alpha>2$. Therefore, also in the case $H(x) = (1-x^2)^2$ we obtain the existence of the critical mass $\mu_c$ coherently with what proven in \eqref{eq:mc2} in the case $m=0$.

\section{Numerical results}\label{sec:numerics}
In this section, we explore the capability of the proposed model \eqref{eq:boltzmann} and of its grazing collision limit \eqref{eq:nonlinFP} to describe certain features related to opinion formation. In particular, as already stated, we focus on the study of conformism within a population and how this mechanism plays a role in shaping ideas in a community. We distinguish in the sequel the case in which opinion is formed and regularly distributed among its possible spectrum, i.e. $x\in[-1,1]$, and we refer to this situation to as the sub-critical case and the condition in which a predominant idea possessed by a given size of population forces other individuals to align in order to conform and avoid strong criticisms from the leading part of the society. We refer to the latter situation as the super-critical case. To highlight the link between the introduced microscopic dynamics and the emerging equilibrium distribution we stick to a Monte Carlo approach \cite{MC}. For the seek of completeness, we mention also  the deterministic numerical methods for similar equations developed in \cite{BMP,CHW,MarPar} and the references therein.
In the first part, we validate the proposed theoretical approach through a body of numerical results. In the second part, we consider several tests highlighting the possibility to reach dynamically a supercritical state in a scenario involving more than one population.

\subsection{Validating the theoretical setting}
\begin{figure*}[h!]\centering
	\includegraphics[width=0.45\textwidth]{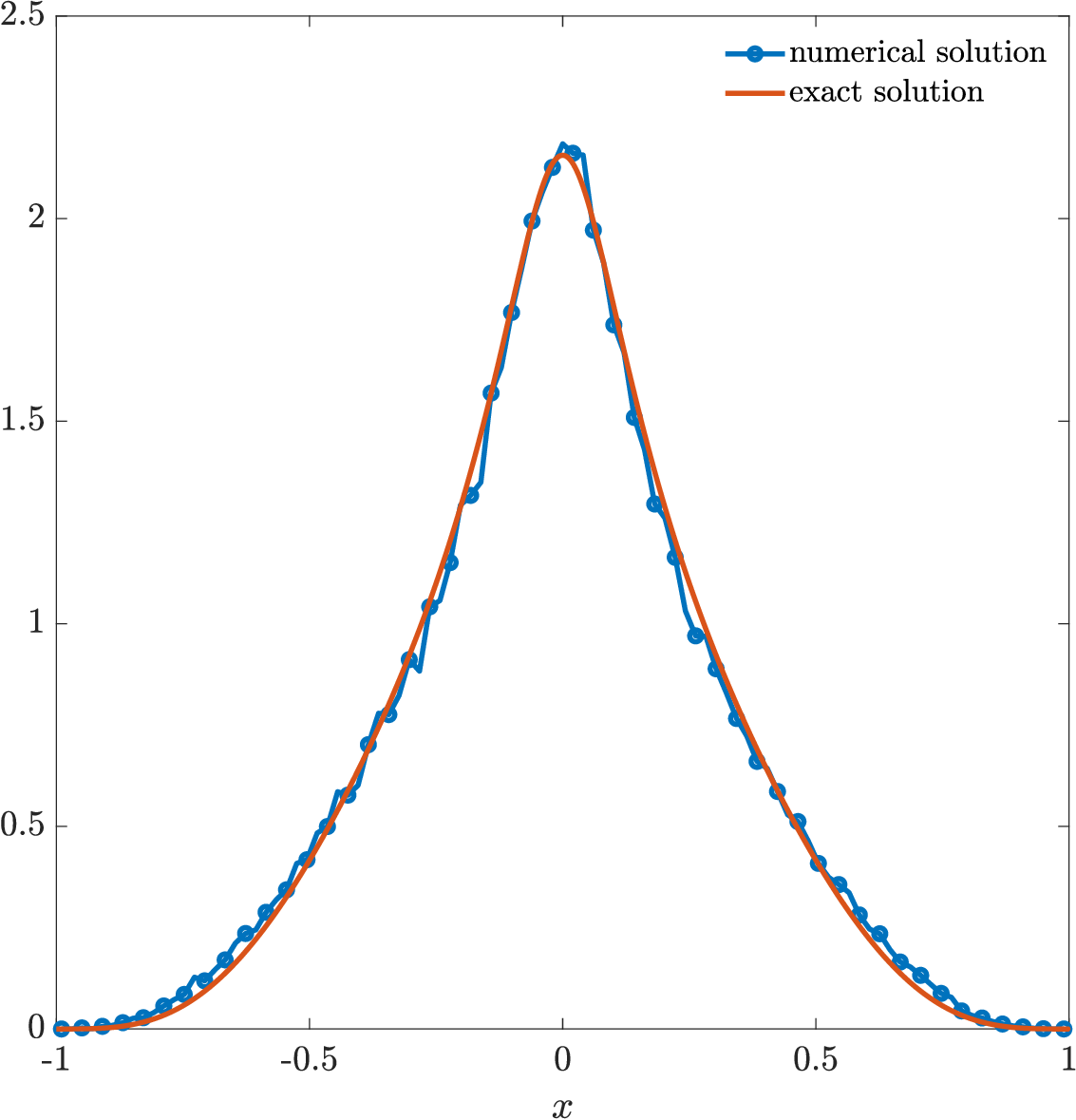} \quad\includegraphics[width=0.45\textwidth]{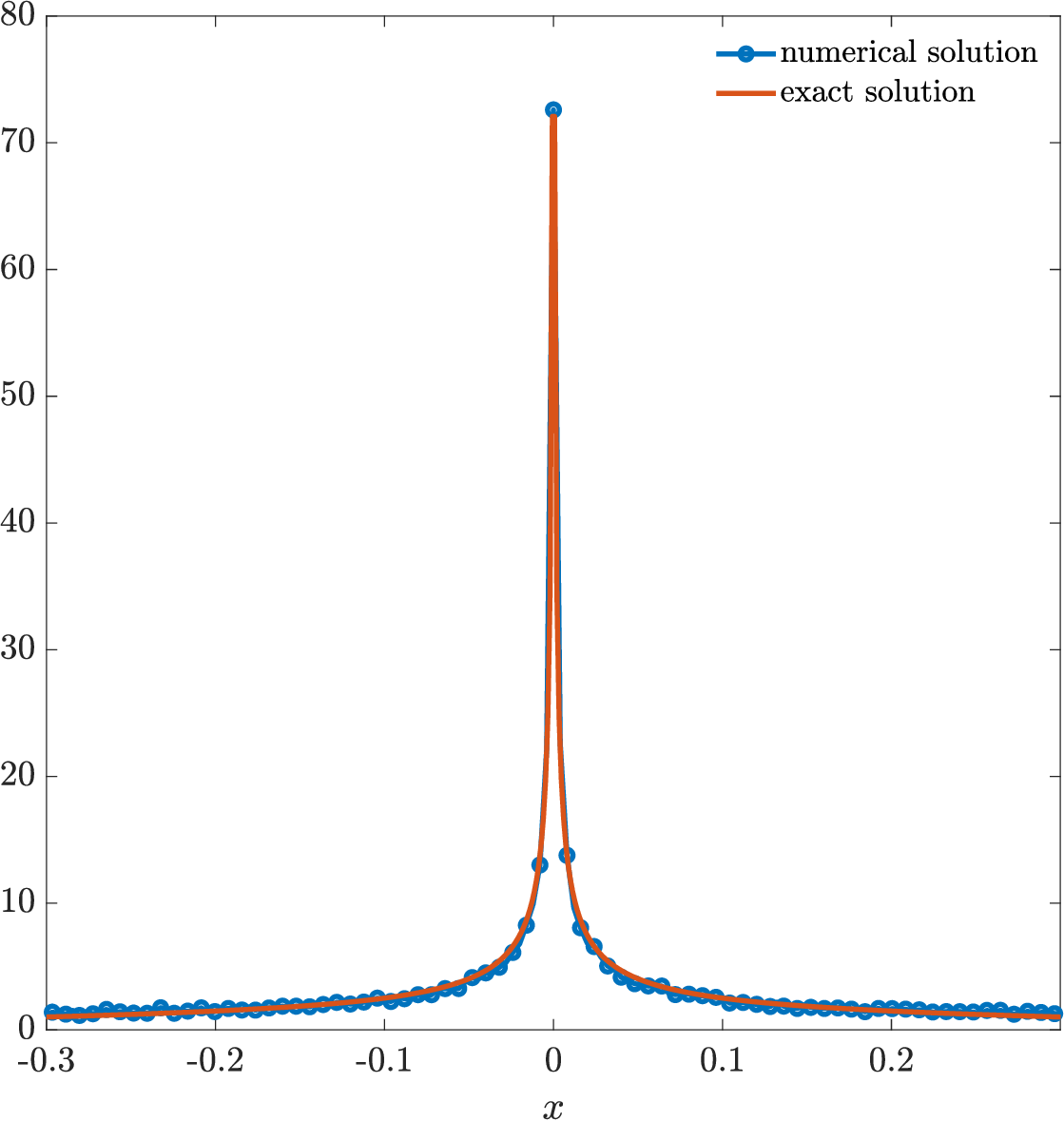}
	\caption{Left sub-critical case obtained with $m = 0$, $\alpha = 3$, $\l = 4$, $\mu = \mu_c/2$. Right super-critical case obtained with $m = 0$, $\alpha = 3$, $\l = 4$, $\mu = \mu_c+0.5$. $H(x)=1-x^2$.}\label{fig:test1_1}
\end{figure*}

In this first set of simulations, we show how the Monte Carlo method applied to the Boltzmann model \eqref{eq:boltzmann} behaves for different values of the modeling and scaling parameters. In particular, we are interested in numerically showing that our method is able to describe the steady state solutions both in the smooth as well as in the blow-up situation and that this steady state corresponds to the one obtained in the previous section for the Fokker-Planck limit. In the following, we first compute the exact stationary solution for various sets of parameters $\alpha, \lambda$ and $m$ and then we compare it with MC. We start from a uniform distribution in $[-1,1]$ and we use $N_s = 10^5$ samples with $H(x)=1-x^2$ and $\epsilon=10^{-3}$. For the simulations shown in Figure \ref{fig:test1_1} we choose $\alpha = 3$, $\lambda = 4$ and $m=0$. The critical mass $\mu_c$ for this choice of parameters reads as in \eqref{eq:massacritica}. Given this value, we compute the stationary states for respectively $\mu = \mu_c/2$ and $\mu = \mu_c+0.5$. In the first case, as expected, the stationary solution is smooth since its mass is below the critical one, while in the second case instead we can clearly see the formation of a singularity in $m=0$, which is very well caught by our Monte Carlo scheme.
\begin{algorithm}[H]
	\caption{Monte Carlo method for the Boltzmann equation \eqref{eq:boltzmann}}\label{nanbu}
	\begin{itemize}
		\item Select a scaling parameter $\var >0$.
		\item Extract $N_s$ samples $x_i^0$ from the initial distribution $f_0(x)$.
		\item Discretize the opinion interval $[-1,1]$ in $N_b$ equal intervals, centered in $y_j$, $j = 1,\dots,N_b$.
		\item The $N_b$ intervals represent the bins of the histogram used to reconstruct the distribution function from the particles knowledge.
		\item Choose a final time $T_f$ and a time step $\Delta t$ such that $\Delta t\leq 1/\var$. This gives $N_t$ the total number of iterations.
	\end{itemize}
	Then
	\begin{algorithmic}[H]
		\For{$n = 0:N_t-1$}	
		\For{$i=1:N_s$}
		\State Find the bin $j(i)$ to which each sample $x_i^n$ belongs and reconstruct $f_{j}^n$.
		\EndFor
		\begin{itemize}
			\item Compute the diffusion coefficient $D_i = \sqrt{\frac{H(x_i)}{1+\beta\left( (1-x_{i}^2) f_{j(x_i)}^n\right)^\alpha}}$ and $\xi$ as in \eqref{eq:update}.
			\item Select $z_i$ samples from the background distribution $g(z)$. 
			\item With probability $\frac{B_\epsilon[f](x_i,t^n)}{\max_{x_i} B_\epsilon[f](x_i,t^n)}$, compute the microscopic change in the agents' opinions as \[x_i^{n+1} =  x_i^n(1-\var\Delta t)+\var\Delta t z_i  + \sqrt{\Delta t D_i} \xi.\]
		\end{itemize}
		\EndFor
	\end{algorithmic}
\end{algorithm}

\begin{figure*}[h!]
	\centering
		\includegraphics[width=0.45\textwidth]{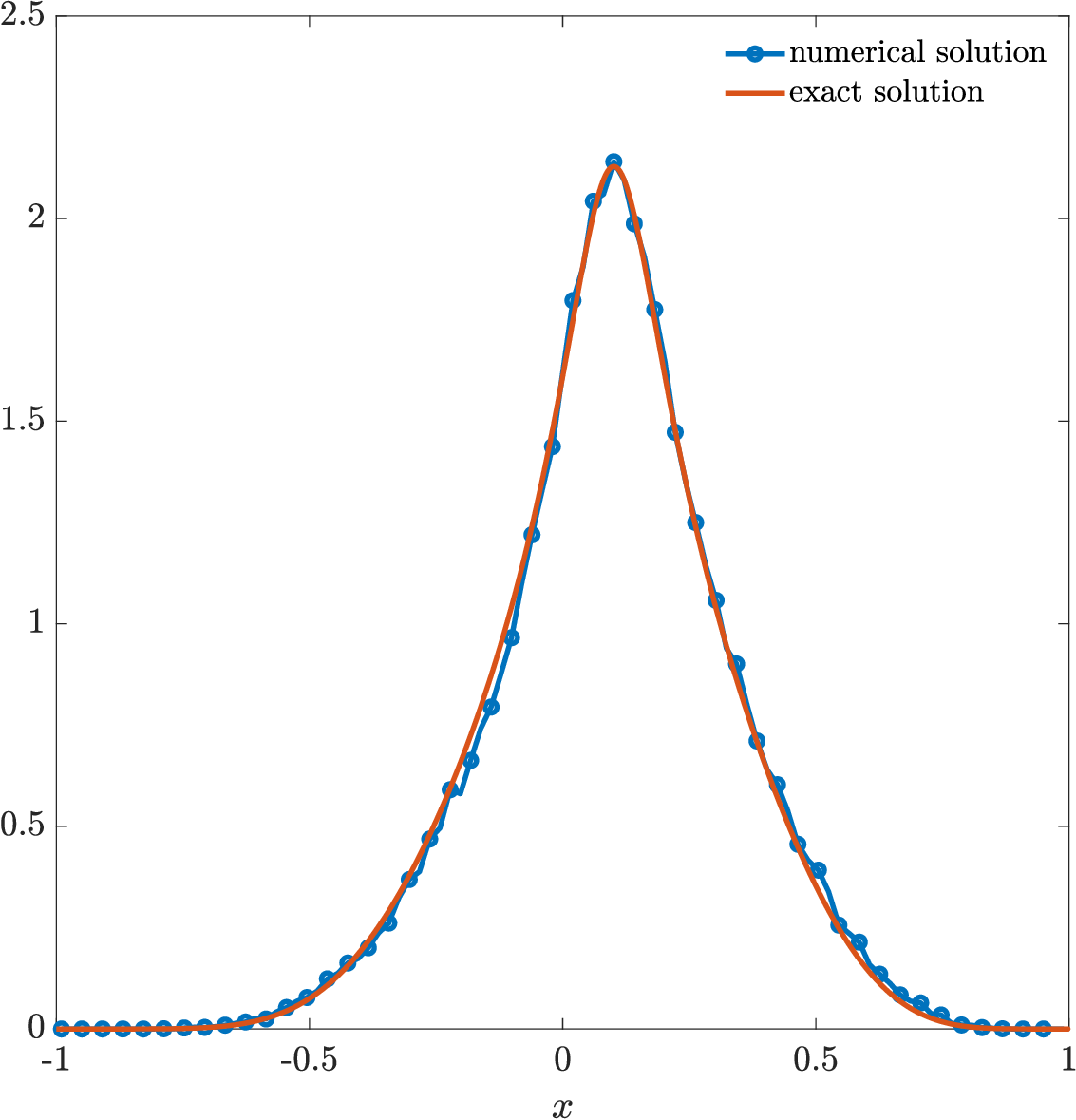} \quad\includegraphics[width=0.45\textwidth]{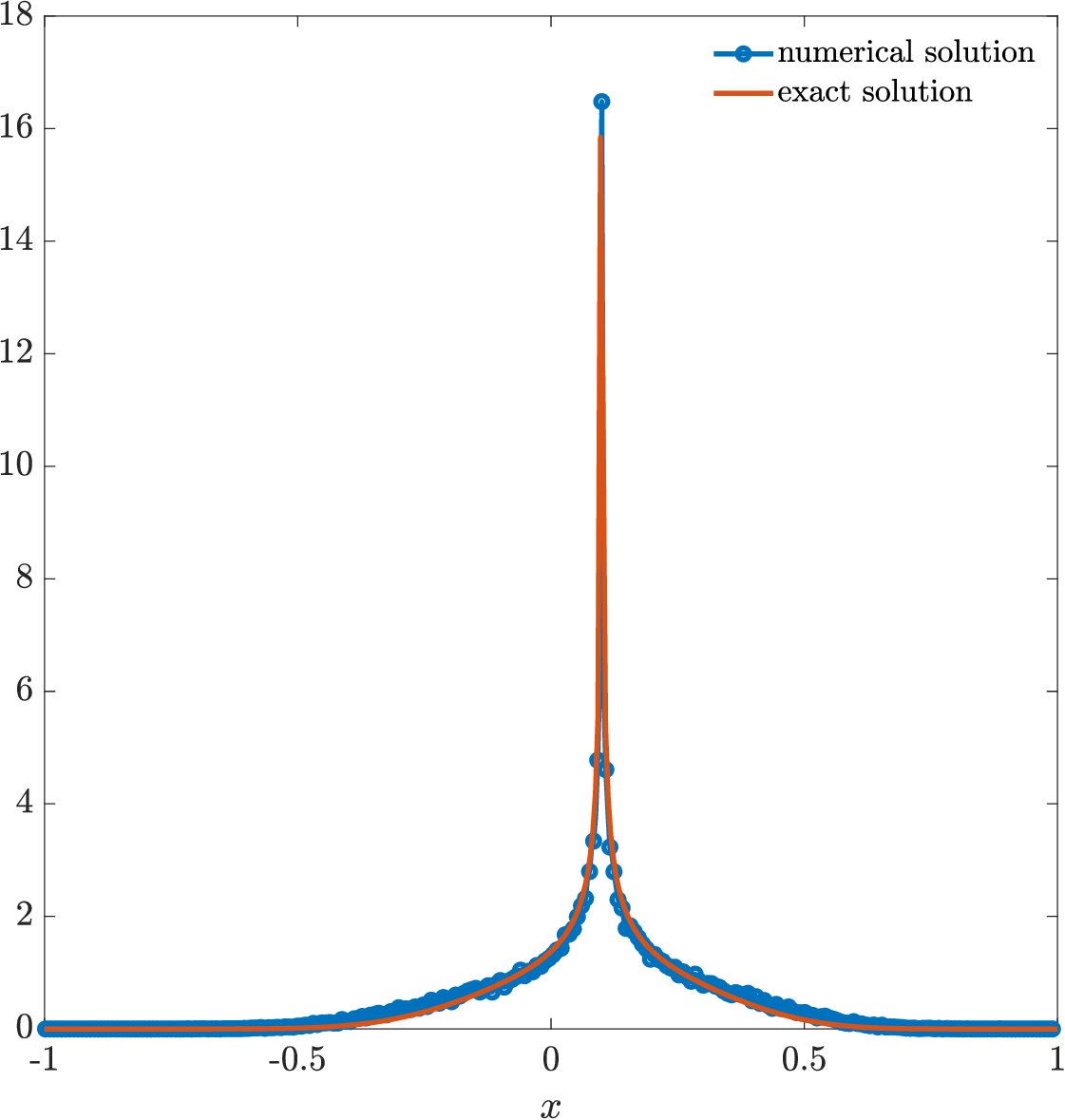}
	\caption{Left sub-critical case obtained with $m = 0.1$, $\alpha = 2.9$, $\l = 7$, $\mu = 1$. Right super-critical case obtained with $m = 0.1$, $\alpha = 4$, $\l = 11$, $\mu = 1$. $H(x)=1-x^2$.}\label{fig:test1_2}
\end{figure*}

\begin{figure*}[h!]
	\centering
	\includegraphics[width=0.45\textwidth]{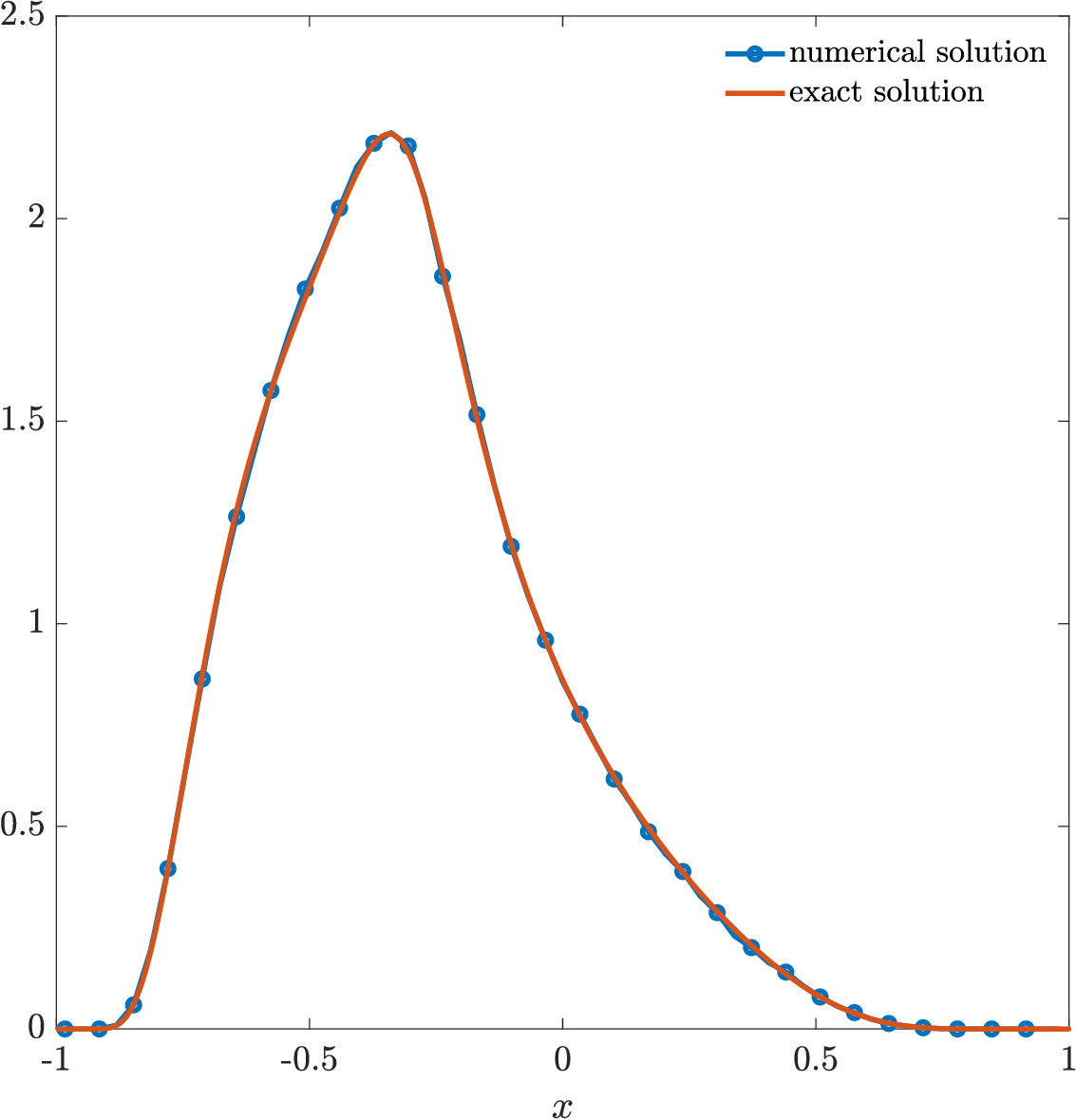}
	\quad
	\includegraphics[width=0.45\textwidth]{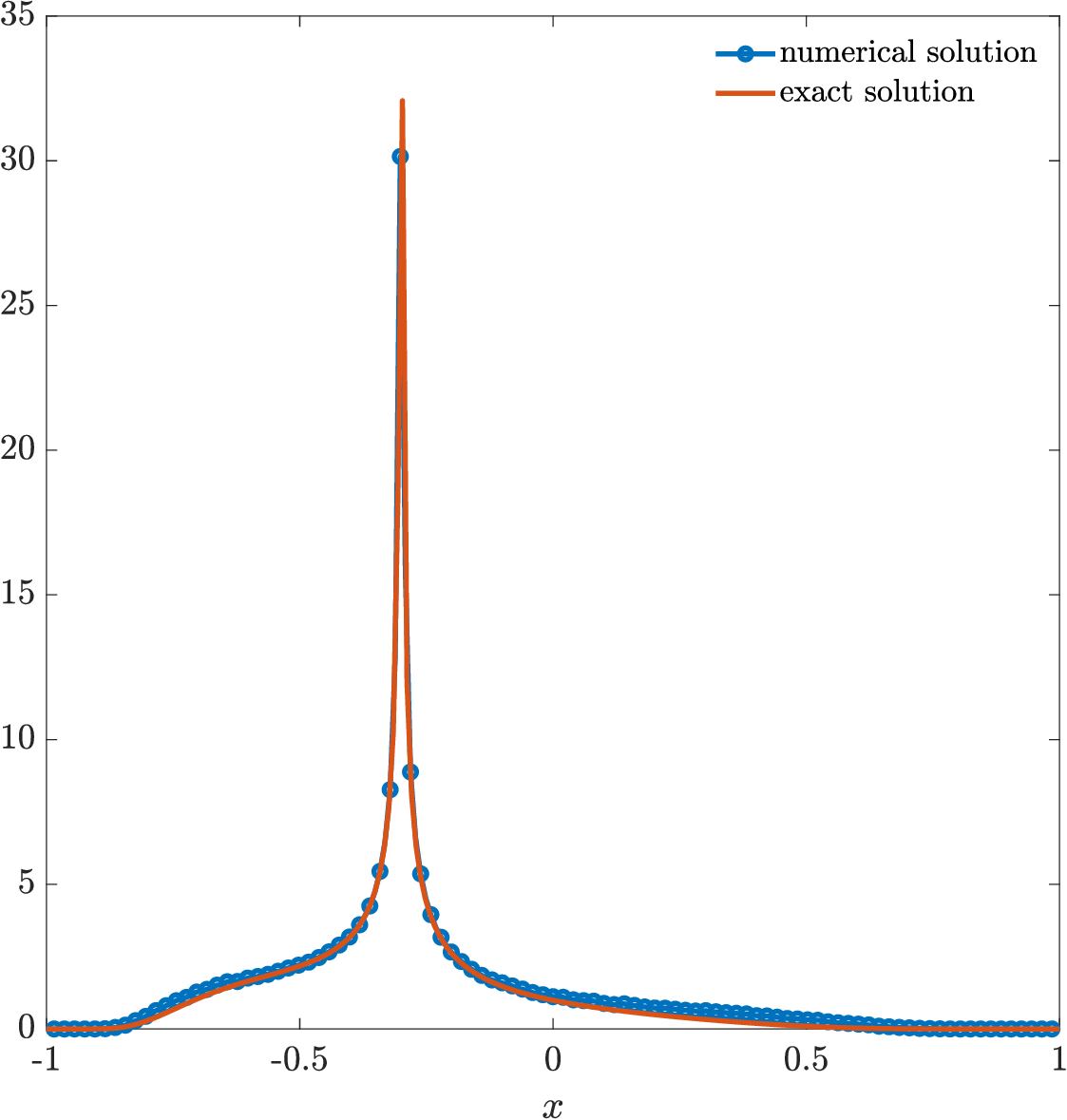}
	\caption{Left sub-critical case obtained with $m = -0.3$, $\alpha = 3$, $\l = 4$, $\mu = \mu_c/2$, right critical case obtained with $m = -0.3$, $\alpha = 3$, $\l = 4$, $\mu = \mu_c$. $H(x)=(1-x^2)^2$.}\label{fig:test1_2b}
\end{figure*}

To underline the importance of the relation existing between the parameters and the possible blow-up around the mean position $m$, we perform another numerical test, now fixing the mass $\mu=1.5$ and showing the different stationary states as we change the values of $\alpha$ and $\lambda$ still keeping $H(x)=1-x^2$. The initial condition is as before the uniform distribution in $[-1,1]$ and, again, we used $N_s = 10^5$ particles with scaling parameter $\epsilon=10^{-3}$. We report the results in Figure \ref{fig:test1_2}. On the left, we can see the stationary state for the choice $\alpha =2.9$,  $\l = 7$, and $m = 0.1$, where the fixed mass $\mu = 1$ is below the critical one, $\mu_c = 2.112$, so that the stationary solution is smooth. On the contrary, on the right, we chose $\alpha = 4$, $\l = 11$, and $m = 0.1$ and for such parameters we observe the super-critical situation: the mass starts to accumulate around $m$. In Figure \ref{fig:test1_2b} we show the case of convergence to equilibrium in the sub and super critical cases where now $H(x)=(1-x^2)^2$. Again, the Monte Carlo method is able to capture the smooth and non smooth equilibrium states with high accuracy.

\begin{figure*}[h!]
	\centering
	\includegraphics[width=0.46\textwidth]{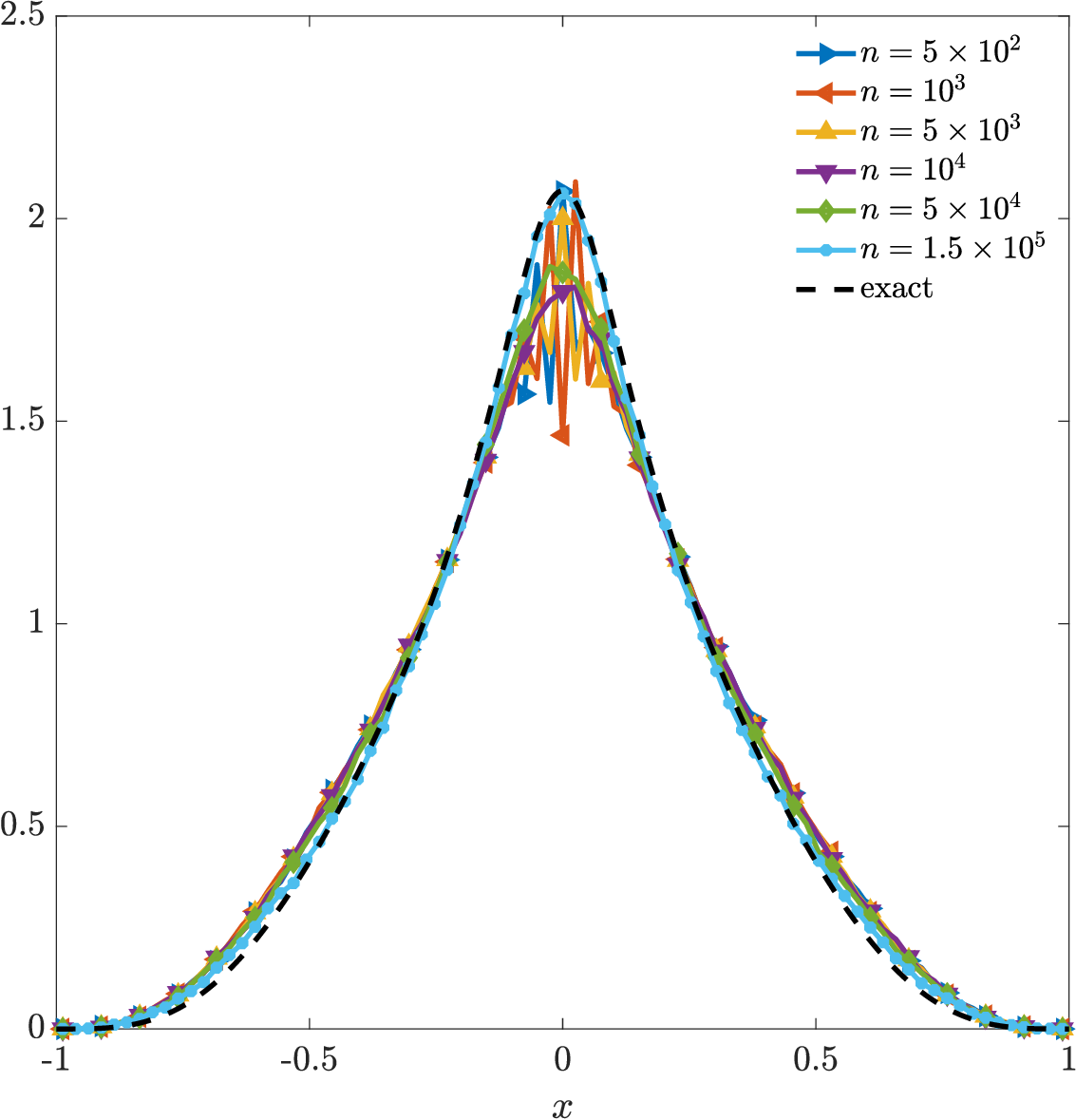} \quad\includegraphics[width=0.45\textwidth]{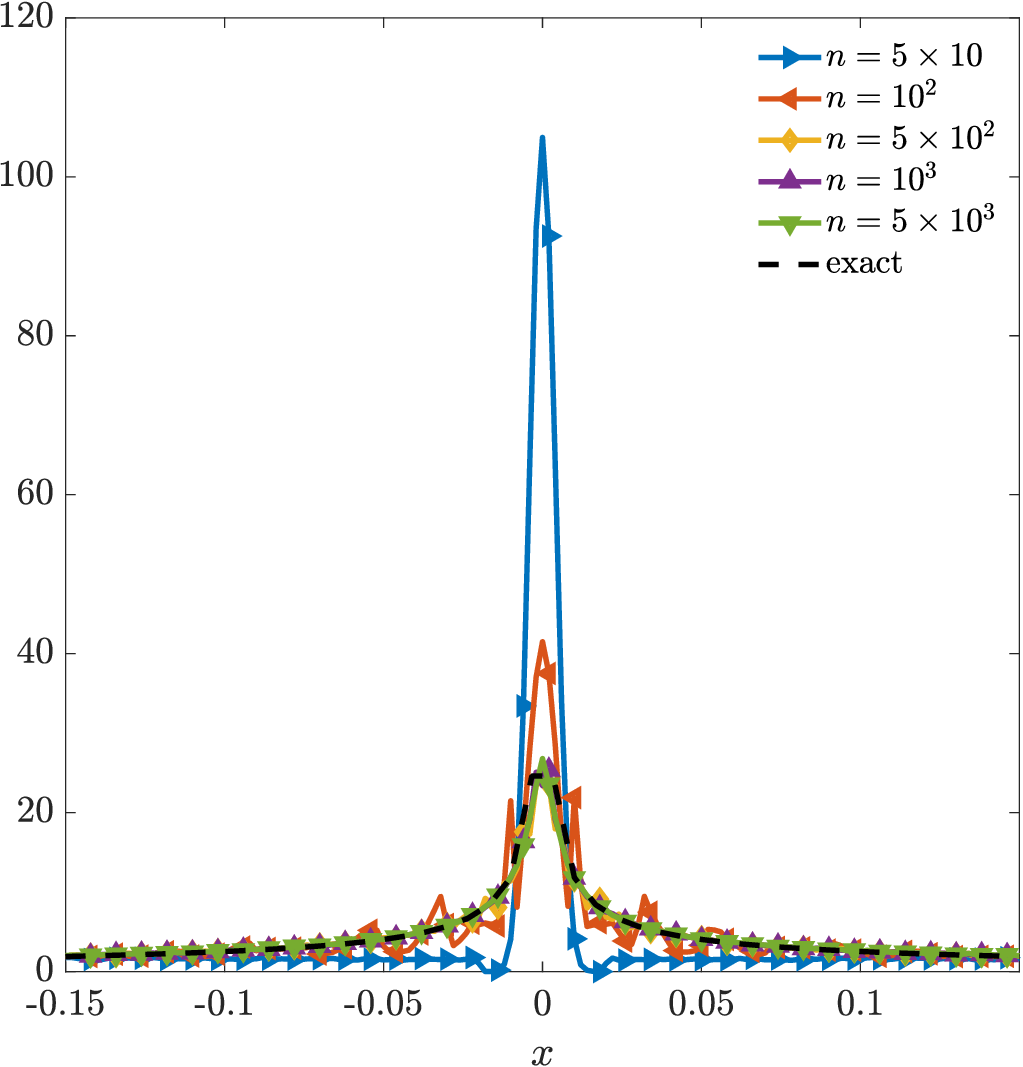}
		\caption{Sensitivity to the number of bins $n$ used in the reconstruction of $f$. Left sub-critical case obtained with $m = 0$, $\alpha = 3$, $\l = 4$, $\mu = 1.278$. Right super-critical case obtained with $m = 0$, $\alpha = 3$, $\l = 4$, $\mu = 2.556$. $H(x)=1-x^2$.}\label{fig:test1_3}
\end{figure*}

In order to show the sensitivity of the Monte Carlo simulations to the mesh size used to reconstruct the density $f$ that is needed to compute the collision frequency and the diffusion term as detailed in algorithm \ref{nanbu}, we perform several simulations with different grid points. To that aim, in Figure \ref{fig:test1_3}, there are shown these results obtained when trying to catch the stationary states with different mesh sizes indexed by $n$, the number of bins necessary to reconstruct $f$, as $\alpha = 3$ and $\lambda = 4$ are fixed, both for the sub-critical case on the left as well as for the super-critical case on the right. We can see how lower values of $n$ may lead to instabilities near the mean value $m$. At the same time, we observe that even higher values of mesh points may not be sufficient to obtain a good representation of the steady states and that the numerical method can fail in representing the solution.

\begin{figure*}[h!]
	\centering
	\includegraphics[width=0.45\textwidth]{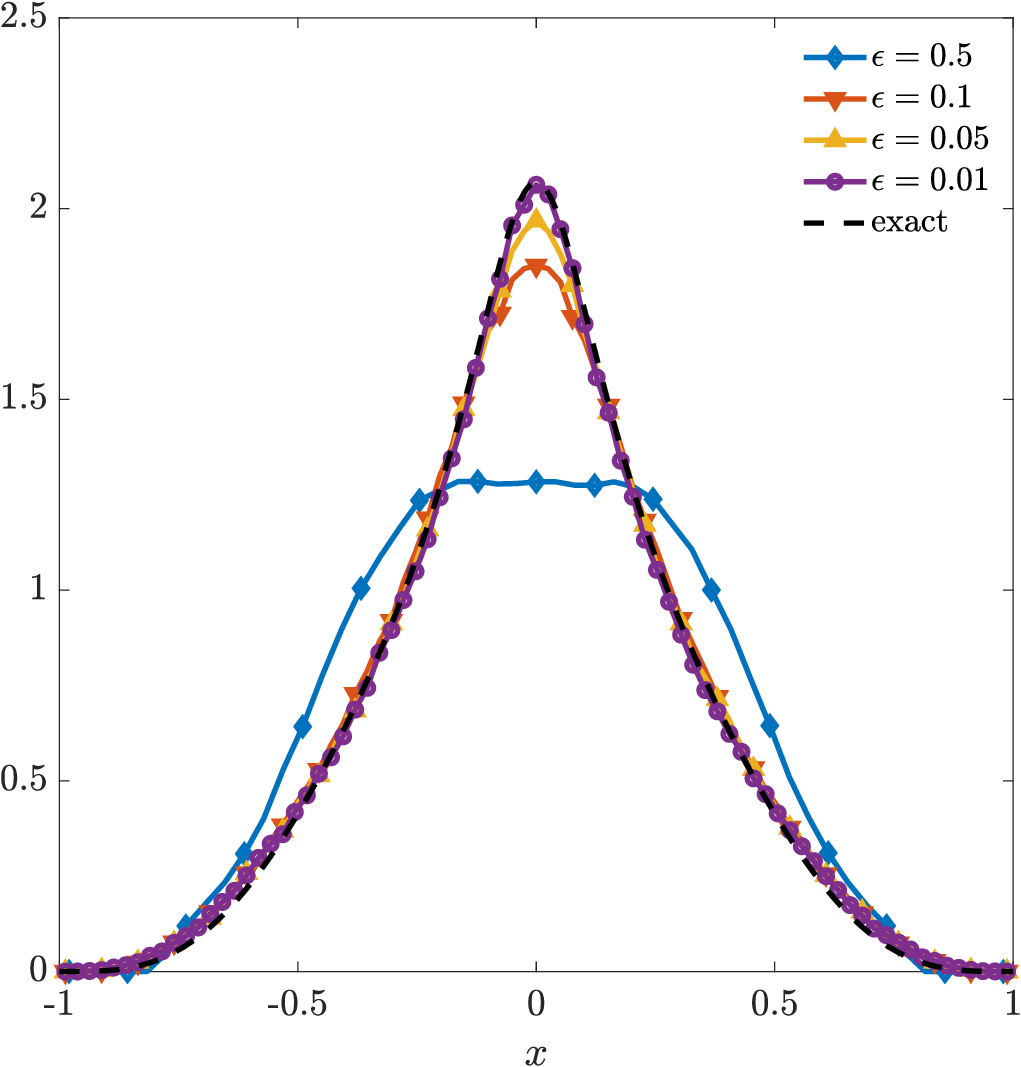}\quad
	\includegraphics[width=0.45\textwidth]{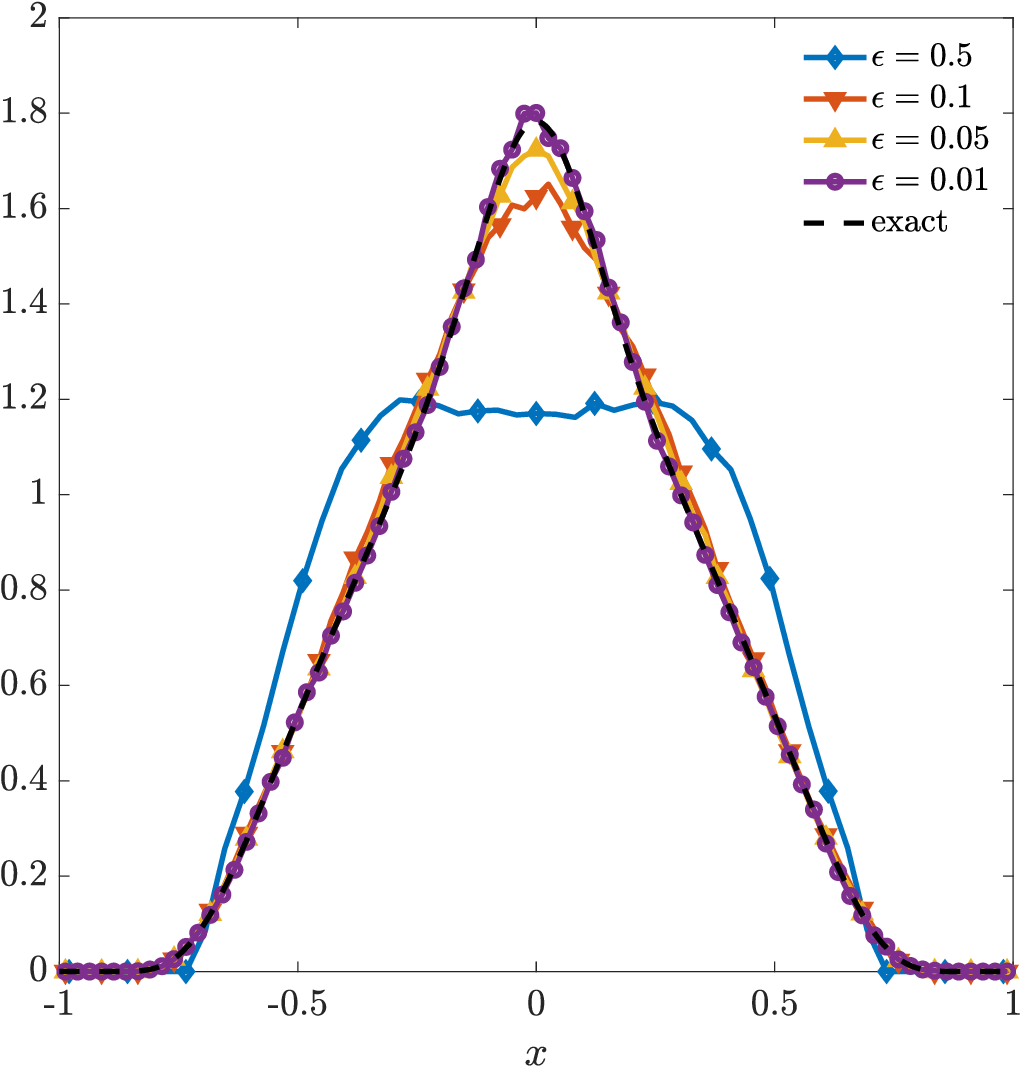}
	\caption{Convergence to the Fokker-Planck model as $\epsilon$ decreases. Left sub-critical case obtained with $m = 0$, $\alpha = 3$, $\l = 4$, $\mu = 1.278$ and $H(x)=1-x^2$. Right sub-critical case obtained with $m = 0$, $\alpha = 3$, $\l = 4$, $\mu = 1.278$ and $H(x)=(1-x^2)^2$.
	}\label{fig:test1_3_}
\end{figure*}

\begin{figure*}[h!]
	\includegraphics[width=0.45\textwidth]{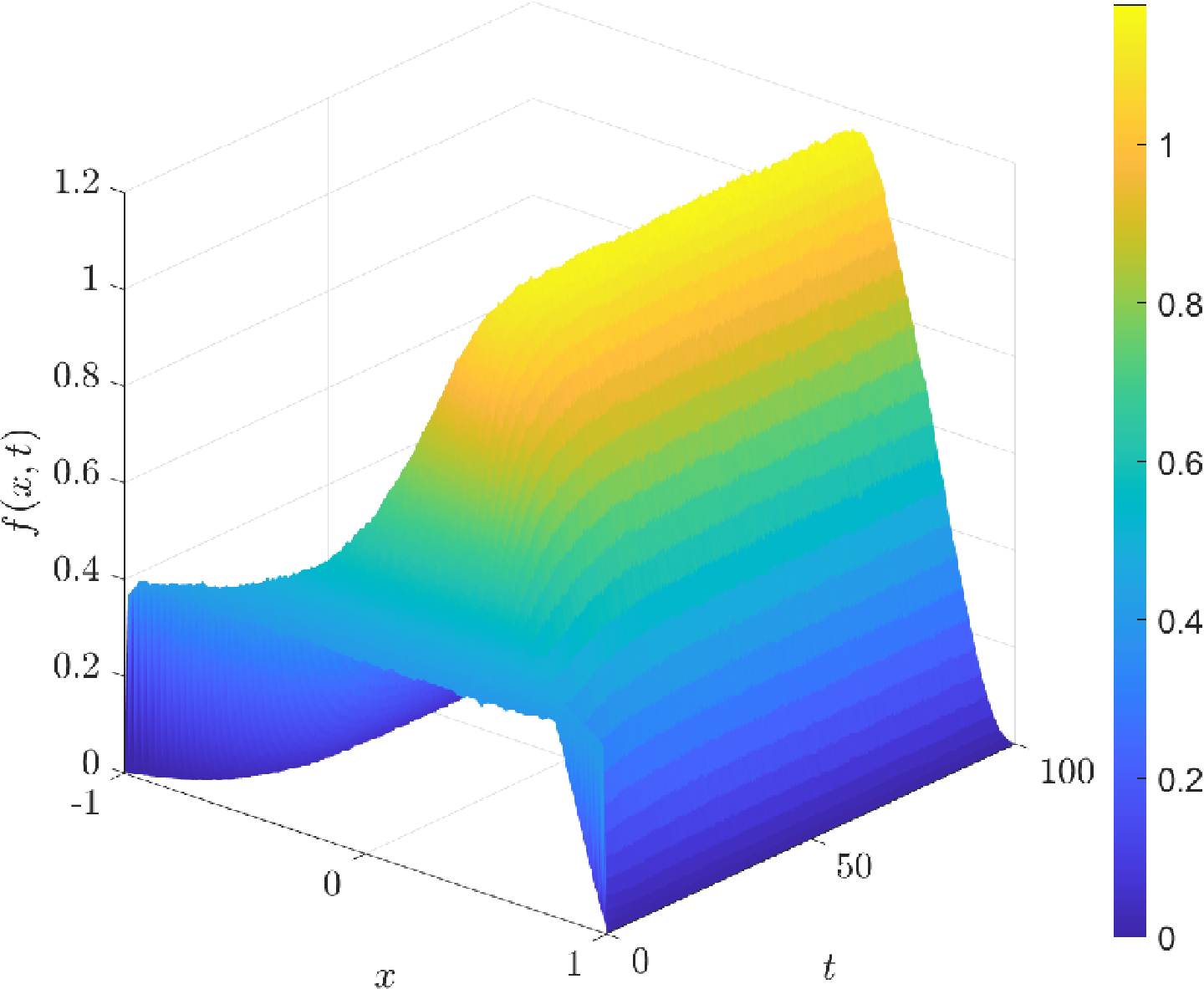} \quad\includegraphics[width=0.45\textwidth]{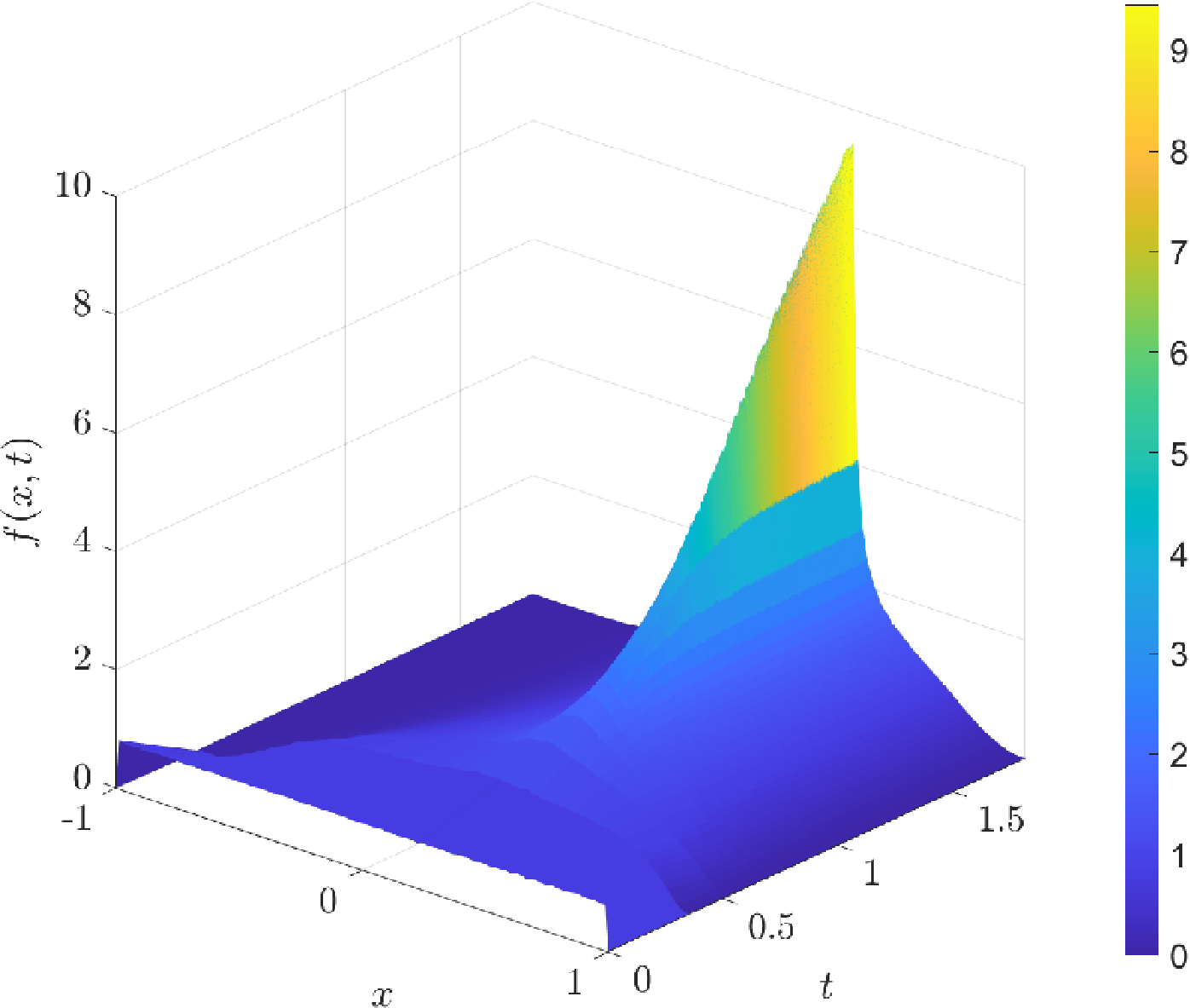}	\caption{Time evolution of the Boltzmann model. Left sub-critical case obtained with $m= 0.3$, $\alpha = 4$, $\l = 5$, $\mu = \mu_c/2$. Right super-critical case obtained with $m= 0.3$, $\alpha = 4$, $\l = 5$ and $\mu = \mu_c$.}\label{fig:test1_3bis}
\end{figure*}

Figure \ref{fig:test1_3_} reports instead the numerical convergence of the Boltzmann model \eqref{eq:boltzmann} to the Fokker-Planck one \eqref{eq:nonlinFP} for different values of the scaling parameter $\epsilon$. We clearly see that, as soon as $\epsilon$ becomes sufficiently small, the Boltzmann model collapses towards the Fokker-Planck solution both for $H(x)=1-x^2$ as well as for $H(x)=(1-x^2)^2$.
On the other hand, when $\epsilon$ is large enough the two dynamics lead to quite different results especially in the smooth case. Figure \ref{fig:test1_3bis} shows for completeness the time evolution in the sub and super critical cases of the Boltzmann model in the Fokker-Planck setting starting from an uniform distribution for a mean opinion $m=0.3$ and $H(x)=1-x^2$.

\begin{figure*}[h!]
\includegraphics[width=0.48\textwidth]{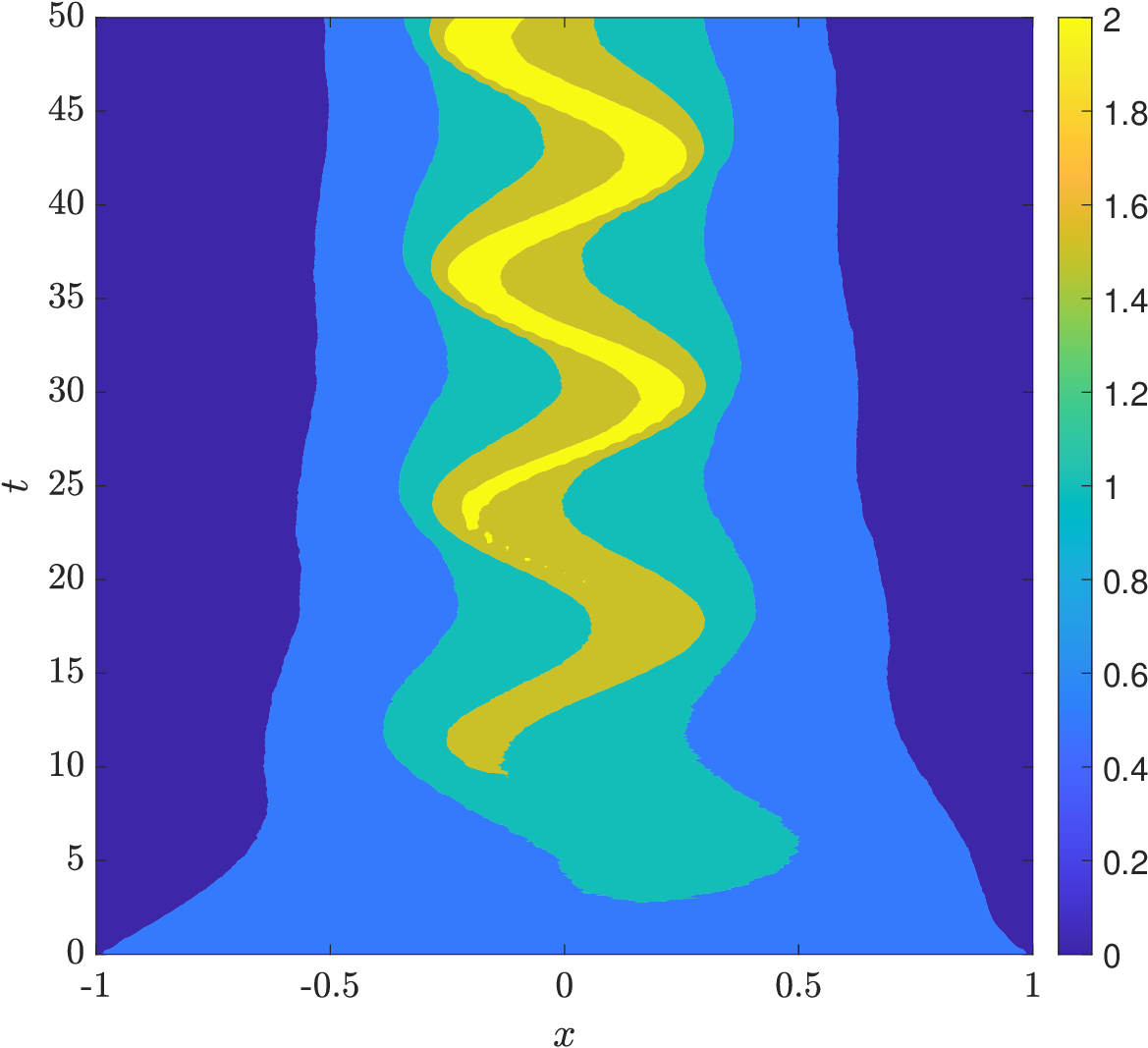} \quad\includegraphics[width=0.5\textwidth]{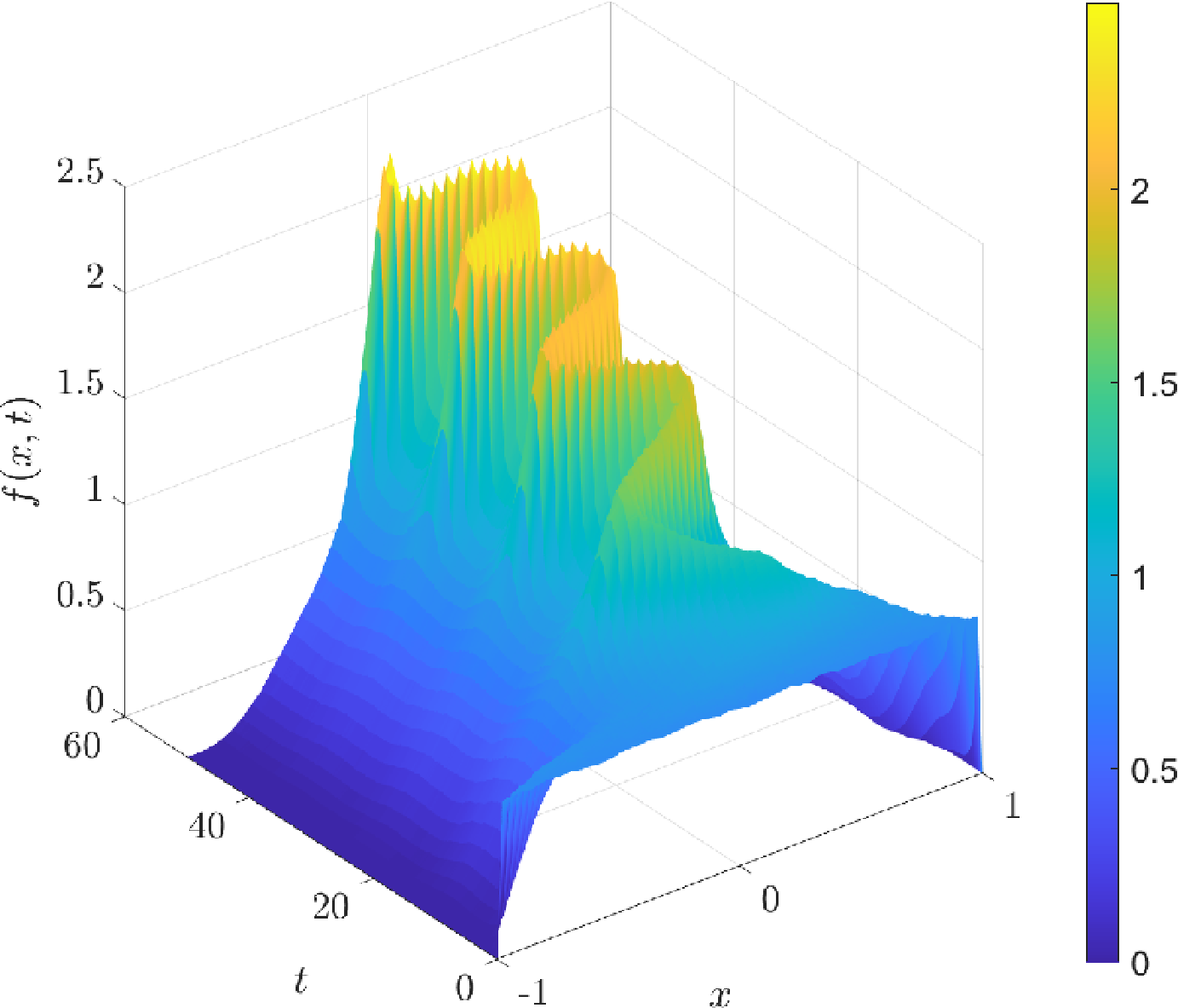}
	\includegraphics[width=0.48\textwidth]{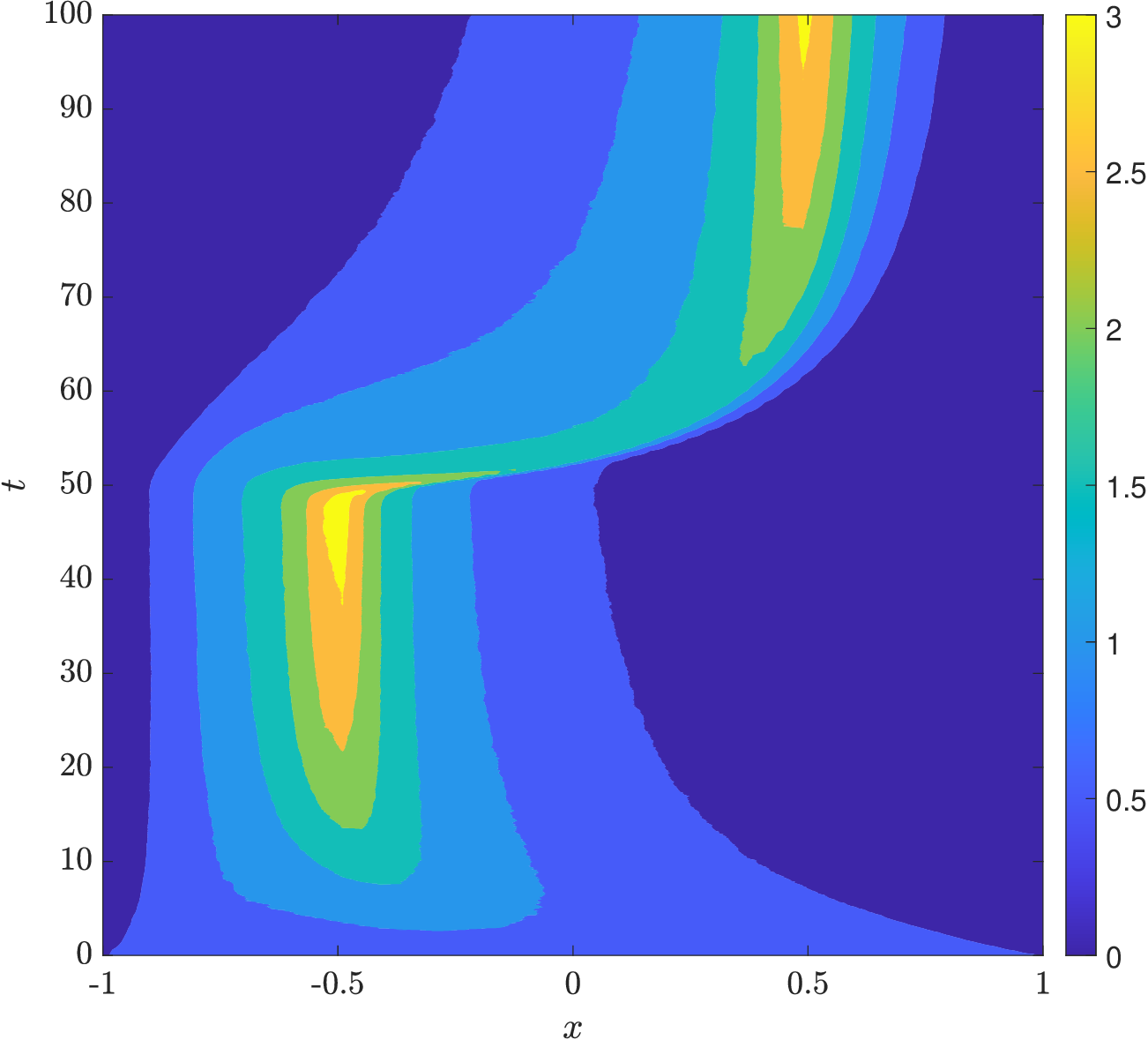} \quad\includegraphics[width=0.5\textwidth]{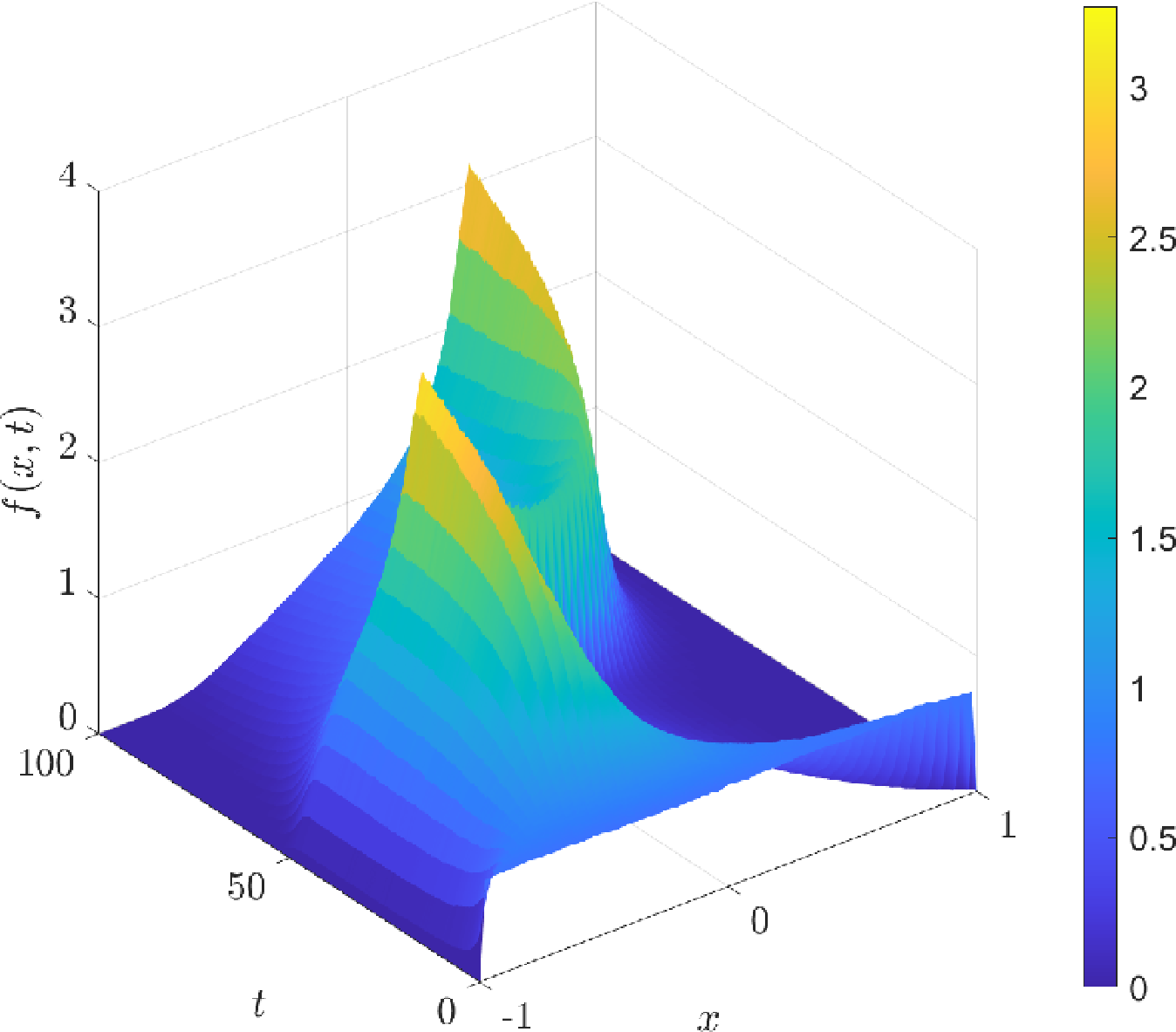}
	\caption{Time evolution of the Boltzmann model with external forces modifying the mean opinion. Top $m(t) = 0.4\sin(0.5t)$, $\alpha = 4$, $\l = 5$, $\mu = 1.5$. Bottom $m(t) = 0.5\mbox{tanh}(t-50)$, $\alpha = 4$, $\l = 5$, $\mu = 1.5$.}\label{fig:test1_4}
\end{figure*}

Finally we consider, to document the qualitative behavior of the model, a situation in which the mean opinion $m$ is not fixed in time while instead it is driven by some external forces. In Figure \ref{fig:test1_4} we show on the top the case $m(t) = 0.4\sin(0.5t)$, while on the bottom we set $m(t) = 0.5\mbox{tanh}(t-50)$. In both simulations we set $\alpha = 4$, $\l = 5$, $\mu = 1.5$ and $H(x)=(1-x^2)$.

\subsection{Interacting populations}\label{2pop}
We now concentrate on the time evolution of opinions between two competing populations of individuals. We suppose that personal ideas may change only as a result of interaction with people belonging to the same group. However, ideas are also shared with all individuals, independently on the group to which the agents belong, and, consequently, one individual may decide to move from one group to the other one when interacting with individuals of the other group with similar ideas. These dynamics are constructed in such a way to mimic the competition between political parties where one party tries to increase its supporters by moving closer to the opinion of individuals belonging to another party in order to capture their interest and tempt them to change faction. 
We suppose then that these two populations are characterized by the time evolution of their internal opinions. The process of opinion formation is regulated by equation \eqref{eq:nonlinFP}, where the variable $x\in[-1,1]$ represents the opinion on a certain topic, with $x=-1$ corresponding to a completely negative opinion on the subject, while $x=1$, on the other hand, signifies a totally positive opinion on the same topic. The function $f_1(t,x)$ now represents the density of opinions at time $t>0$ of the first population while $f_2(t,x)$ of the second one. We suppose that, at time $t=0$, their densities $f_{1,0}$ and $f_{2,0}$ are given by
{
\[
f_1(x,0) = 
\begin{cases}
C_1 \exp\left\{-\frac{1}{2}\frac{(x+0.1)^2}{0.05^2}\right\} & x \in [-1,1] \\
0 & \textrm{elsewhere},
\end{cases}
\]
and
\[
f_2(x,0) = 
\begin{cases}
C_2 \exp\left\{-\frac{1}{2}\frac{(x-0.2)^2}{0.1^2}\right\} & x \in [-1,1] \\
0 & \textrm{elsewhere}, 
\end{cases}
\]
}
with $C_1,\,C_2$ constants such that 
\[
\int_{\mathcal I} f_1(x,0)\mathrm{d}x = \rho_1(t=0)\quad \mbox{and} \quad \int_{\mathcal I} f_2(x,0)\mathrm{d}x = \rho_2(t=0),
\]
where $\rho_i(t)$, $i=1,2$, indicates the total mass of population $i$ at time $t$ and for each $t \geq 0$,  $\rho_1(t) + \rho_2(t) = m_{tot}$, i.e. the total number of people, remains constant over time.

\begin{figure*}
	\centering
	\includegraphics[width=0.3\textwidth]{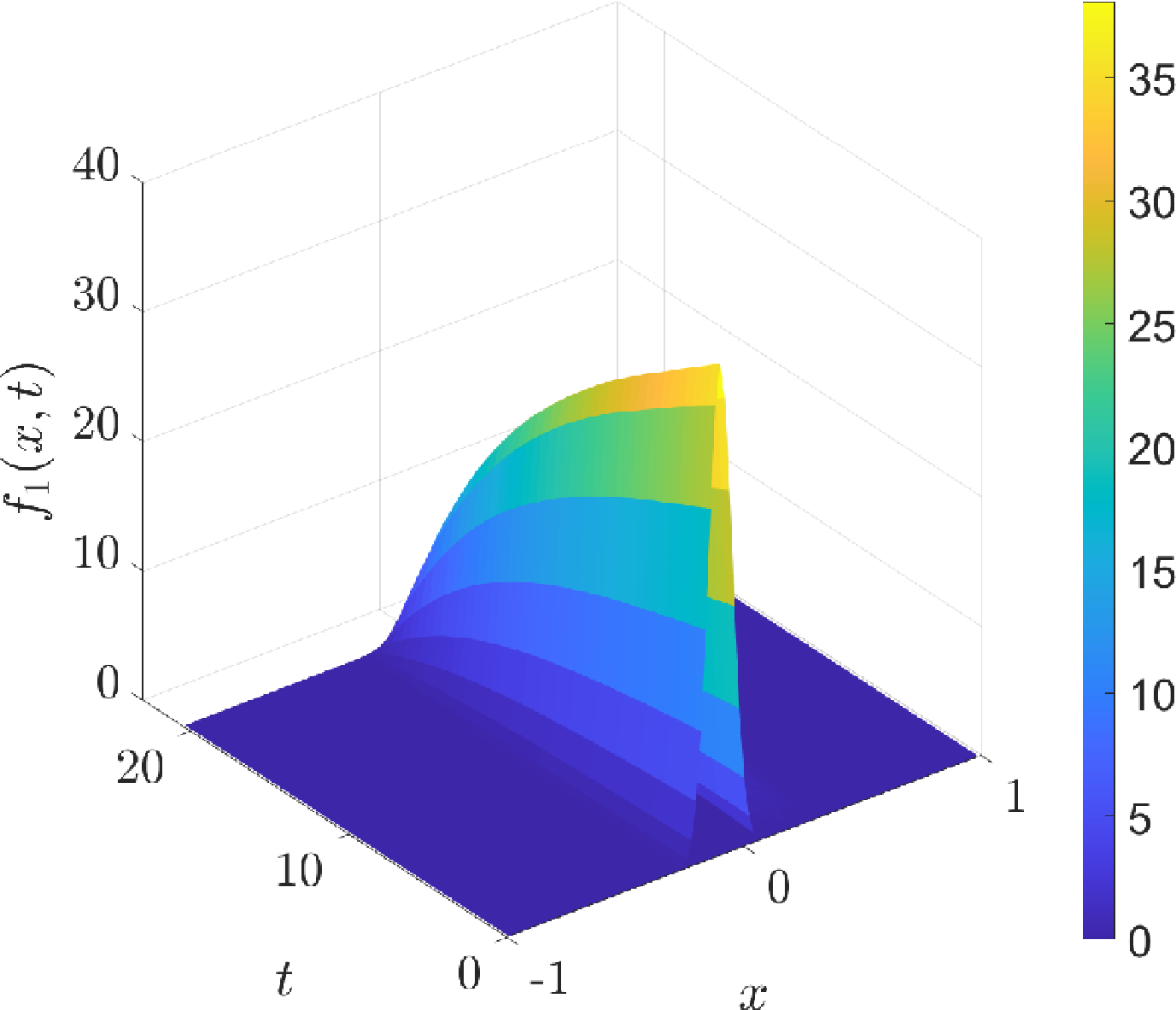}  \includegraphics[width=0.3\textwidth]{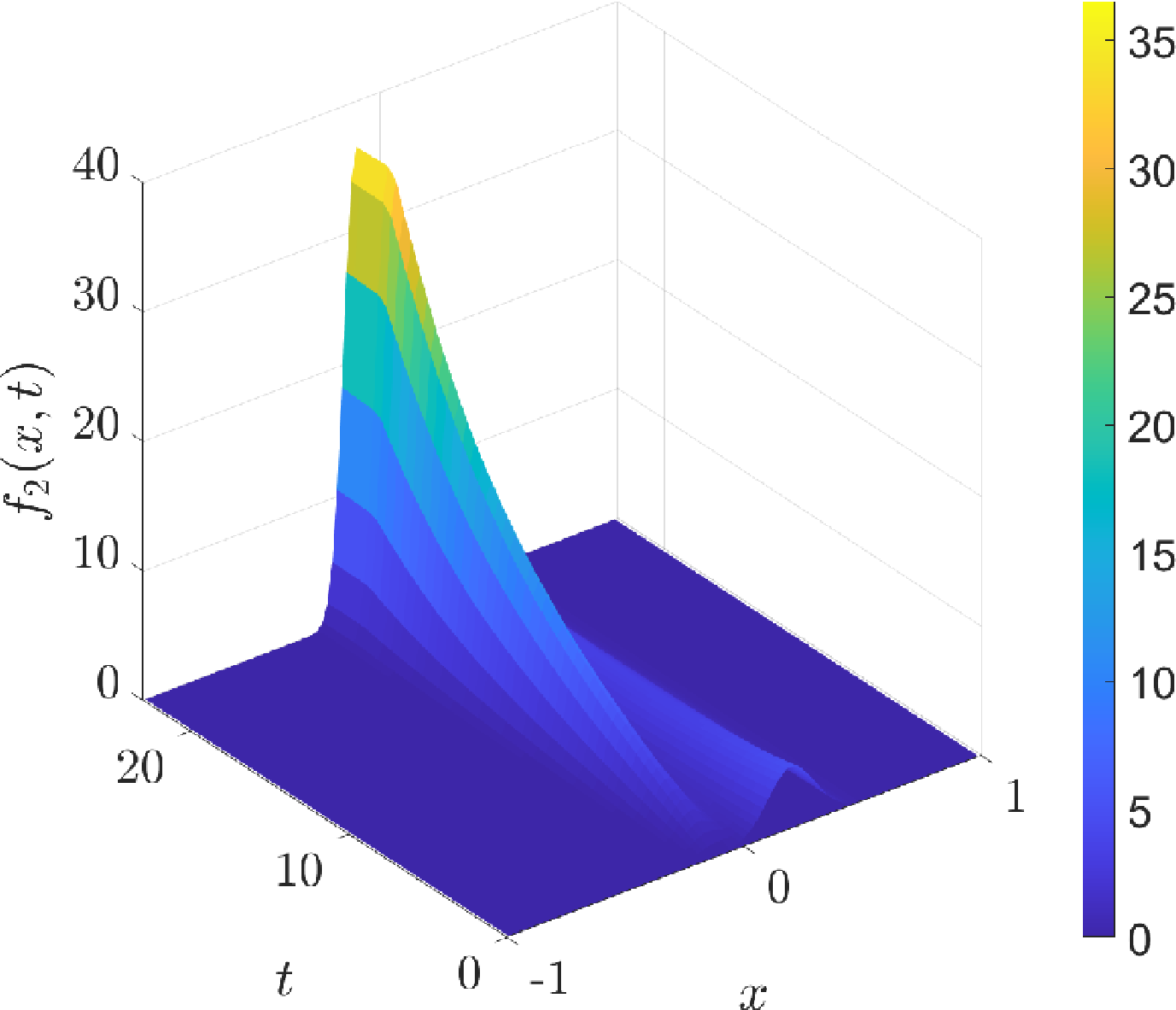}\includegraphics[width=0.3\textwidth]{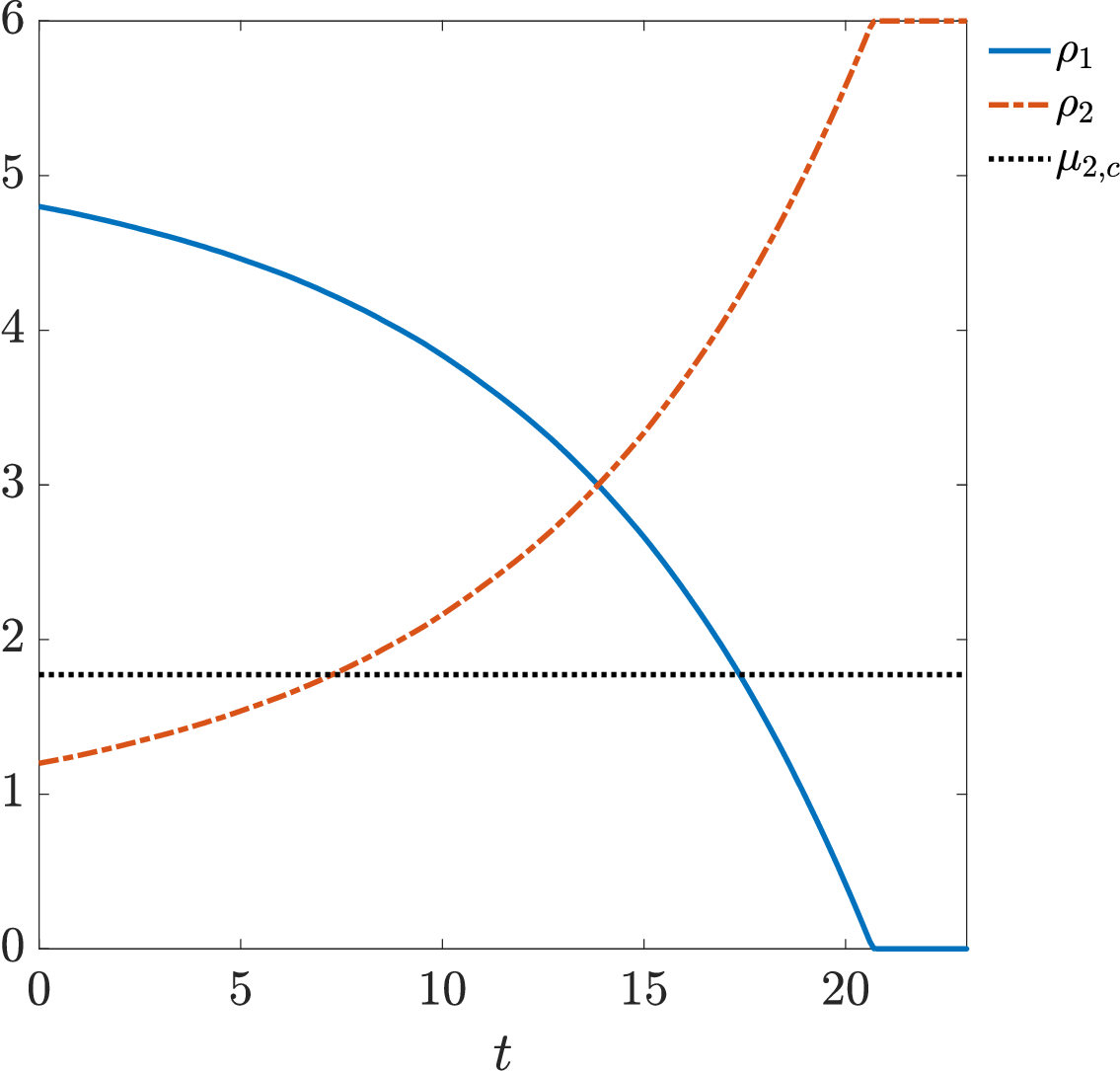}
	\caption{{Time evolution of $f_1$ (left) and $f_2$ (right) for the case $\tau = 10^{2}$.}}
	\label{fig:test_2pop1}
\end{figure*}

The competing populations model is then driven by the following system of equation
\begin{equation}\label{eq:system_inter_pop}
	\begin{cases}
		\partial_t f_1(t,x) = -K_1(f_1,f_2) + K_2(f_2,f_1) + \frac{1}{\tau} Q^1_{BE}(f_1,f_1),\\
		\partial_t f_2(t,x) = K_1(f_1,f_2) - K_2(f_2,f_1)  + \frac{1}{\tau} Q^2_{BE}(f_2,f_2),
	\end{cases}
\end{equation}
where the operators $K_i$, $i=1,2$, regulate the exchange between the two groups  and $Q^i_{BE}$ is the Fokker-Planck operator modeling the shape of political positions. The interaction between the two groups is defined as
\begin{equations}\label{eq:interaction2pop}
	K_1(f_1,f_2) = &\frac{f_1(t,x)}{\rho_1(t)m_{tot}} \int_{-1}^1 \kappa_1(x,x_*)f_2(t,x_*) \mathrm{d}x_*, \\
	K_2(f_2,f_1) = &\frac{f_2(t,x)}{\rho_2(t)m_{tot}} \int_{-1}^1 \kappa_2(x,x_*)f_1(t,x_*) \mathrm{d}x_*, \\
\end{equations}
with
\begin{equation}\label{eq:kernel2pop}
	\kappa_{1,2} (x,x_*) = \chi \left(\left|x-x_* \right| \leq \Delta_{1,2}\right)\gamma_{1,2}.
\end{equation}
The constants $0 \leq \gamma_{1,2} \leq 1$ represent the transition rates between populations and the presence of the indicator function $\chi \left(\left|x-x_* \right| \leq \Delta_{1,2}\right)$ models the fact that the transition probability is strictly positive only if the opinions between two individual belonging to different groups is close enough, i.e. they have a distance less or equal than $\Delta_{1,2} \in [0,2]$. This latter being considered a characteristic parameter of each population. Finally, $\tau>0$ plays the role of a time scaling parameter and its value characterizes the different velocities at which the two above described mechanisms take place.

\begin{figure*}
	\centering
	\includegraphics[width=0.3\textwidth]{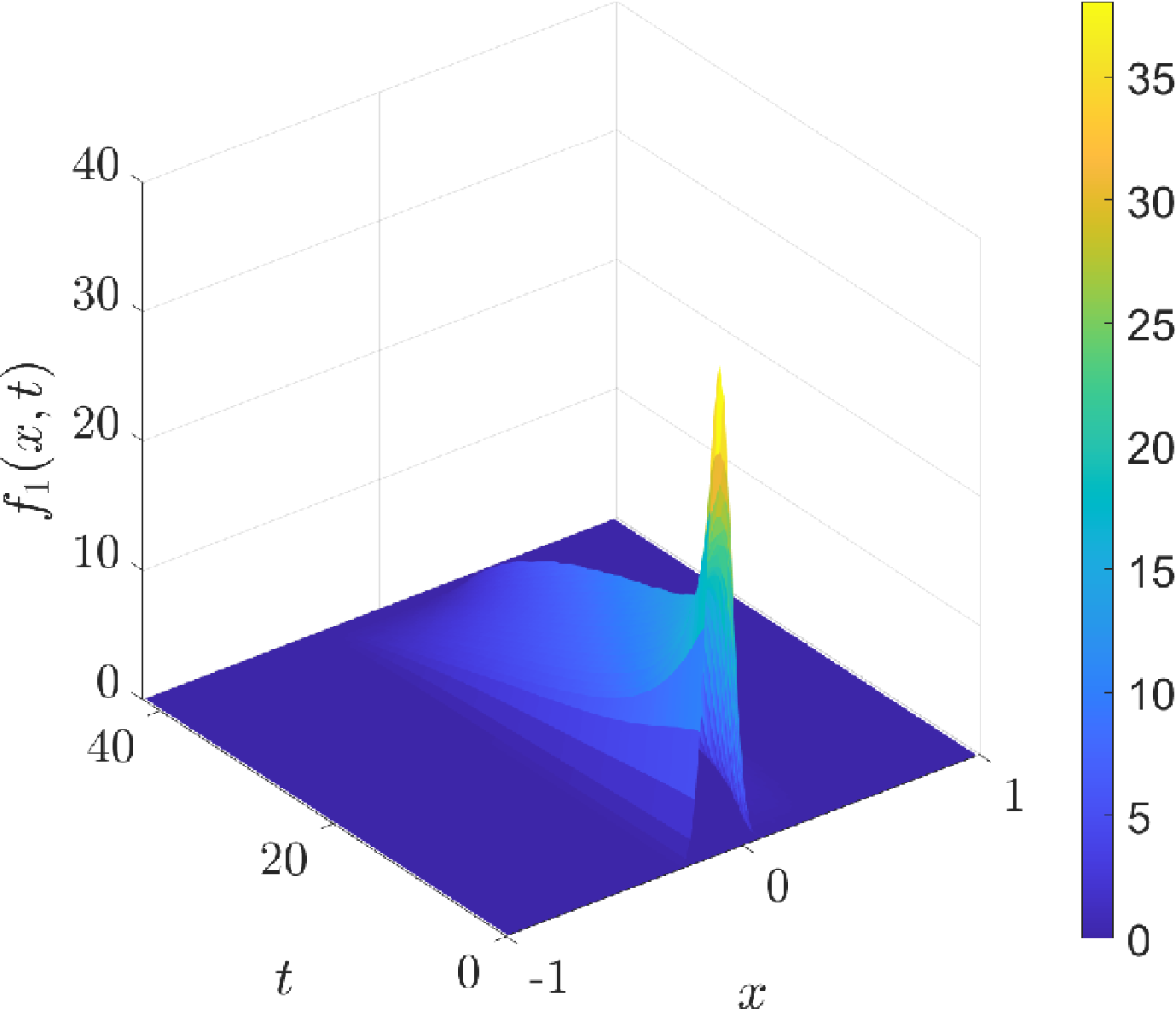}  						\includegraphics[width=0.3\textwidth]{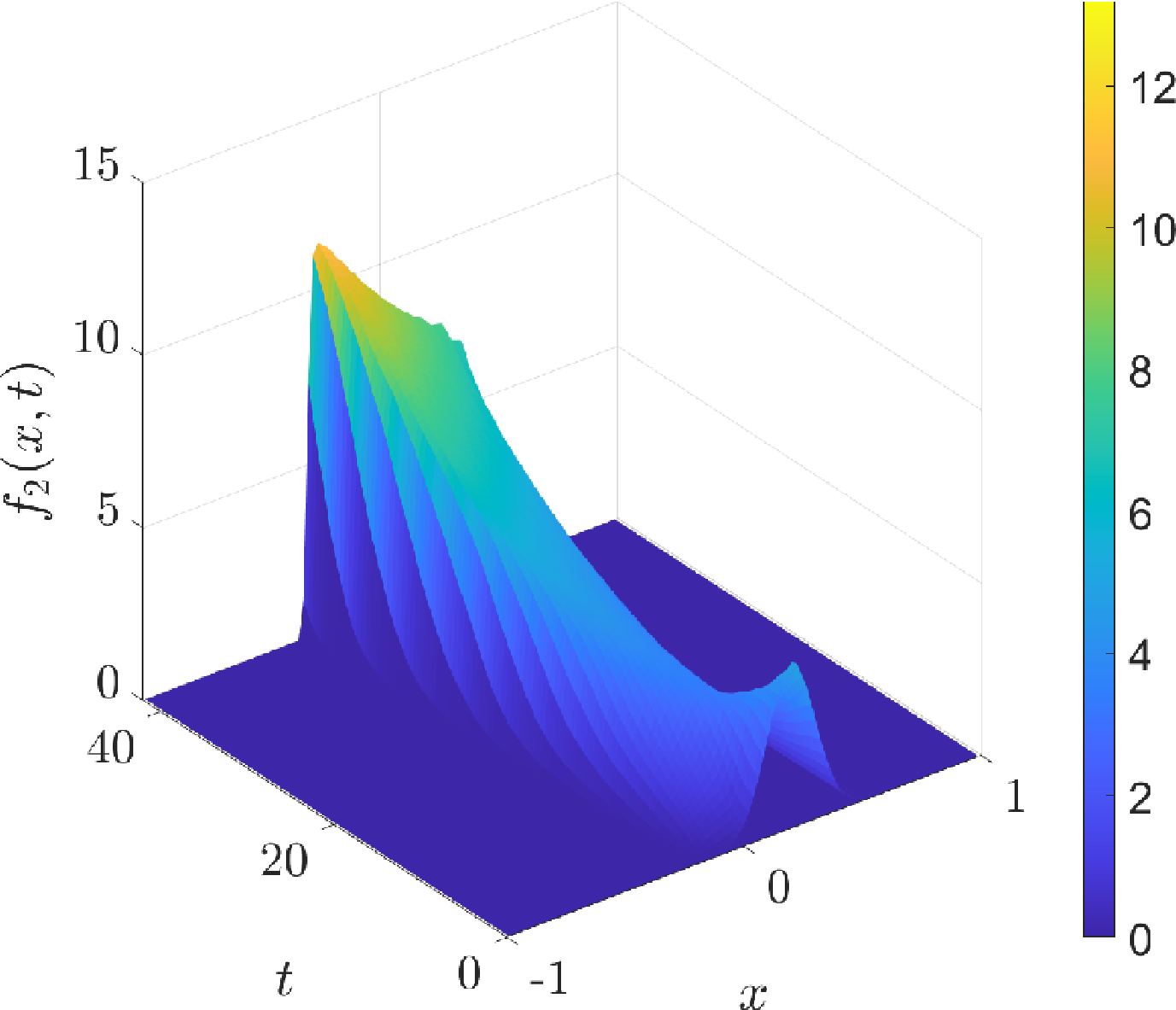}
	\includegraphics[width=0.3\textwidth]{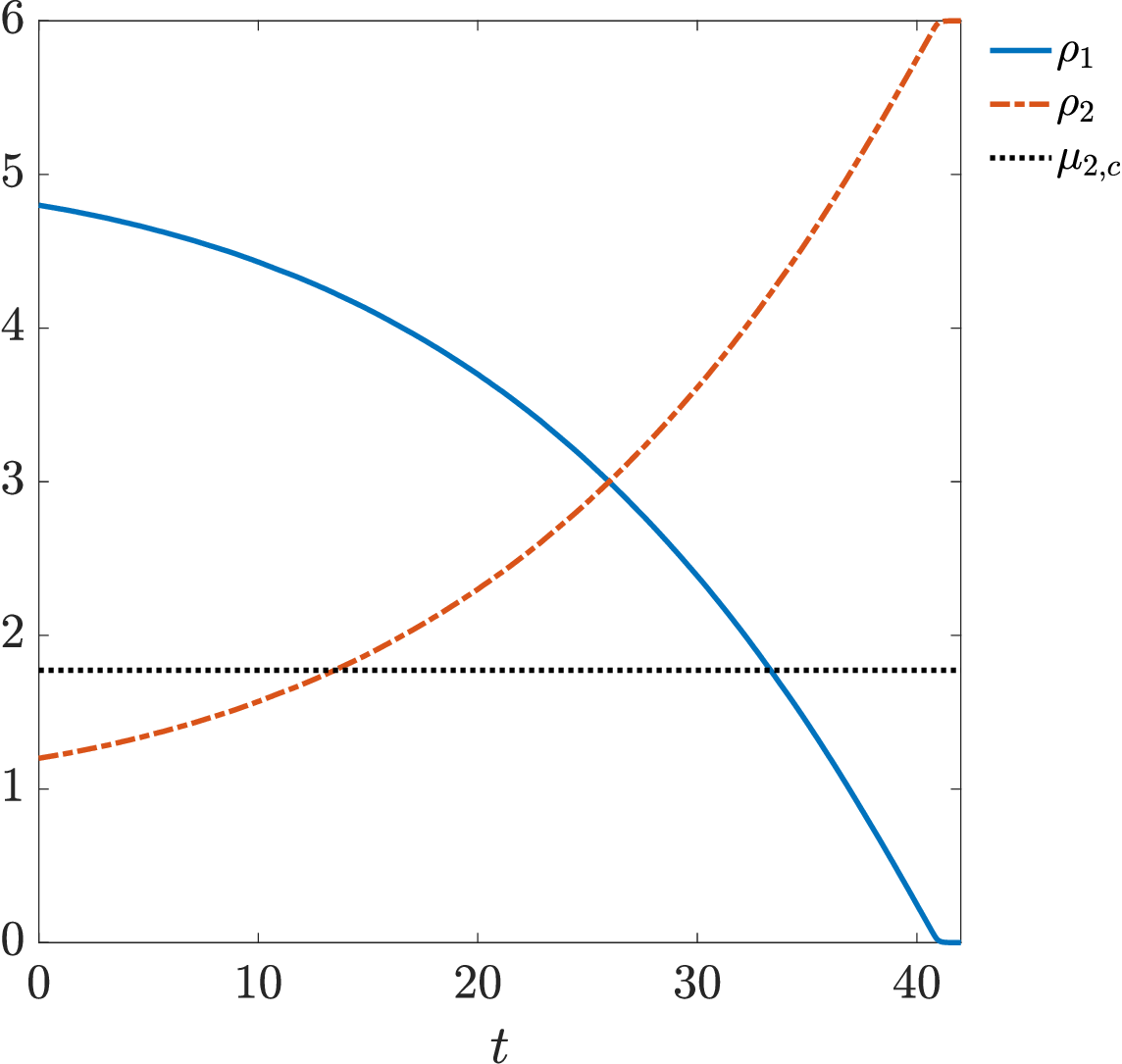}	
	\caption{{Time evolution of $f_1$ (left) and $f_2$ (right) for the case  $\tau = 1$.}}
	\label{fig:test_2pop2}
\end{figure*}

In this depicted context, we perform now three different simulations, for respectively $\tau = 10^2,1,10^{-2}$, each of them with $\alpha_1 = \alpha_2 = 4$, $\lambda_1 = 6$, $\lambda_2 = 5$, $\Delta_1 = \Delta_2 = 0.4$, $\gamma_1 = 0.7$ and $\gamma_2 = 0.1$. The accumulation points, i.e. the mean opinions, are $m_1 = 0.5$ and $m_2 = -0.5$ for the two groups. The initial total mass of the system is set $\mu_{tot} = 6$,  with $\rho_1(0) = 0.8\mu_{tot}$ and $\rho_2(0) = 0.2\mu_{tot}$, meaning that at time $t=0$ both the densities $f_1,\, f_2$ are below the critical mass for their respective collision operator $\Q^i_{BE}$, $i=1,2$. 

\begin{figure*}
	\centering
	\includegraphics[width=0.4\textwidth]{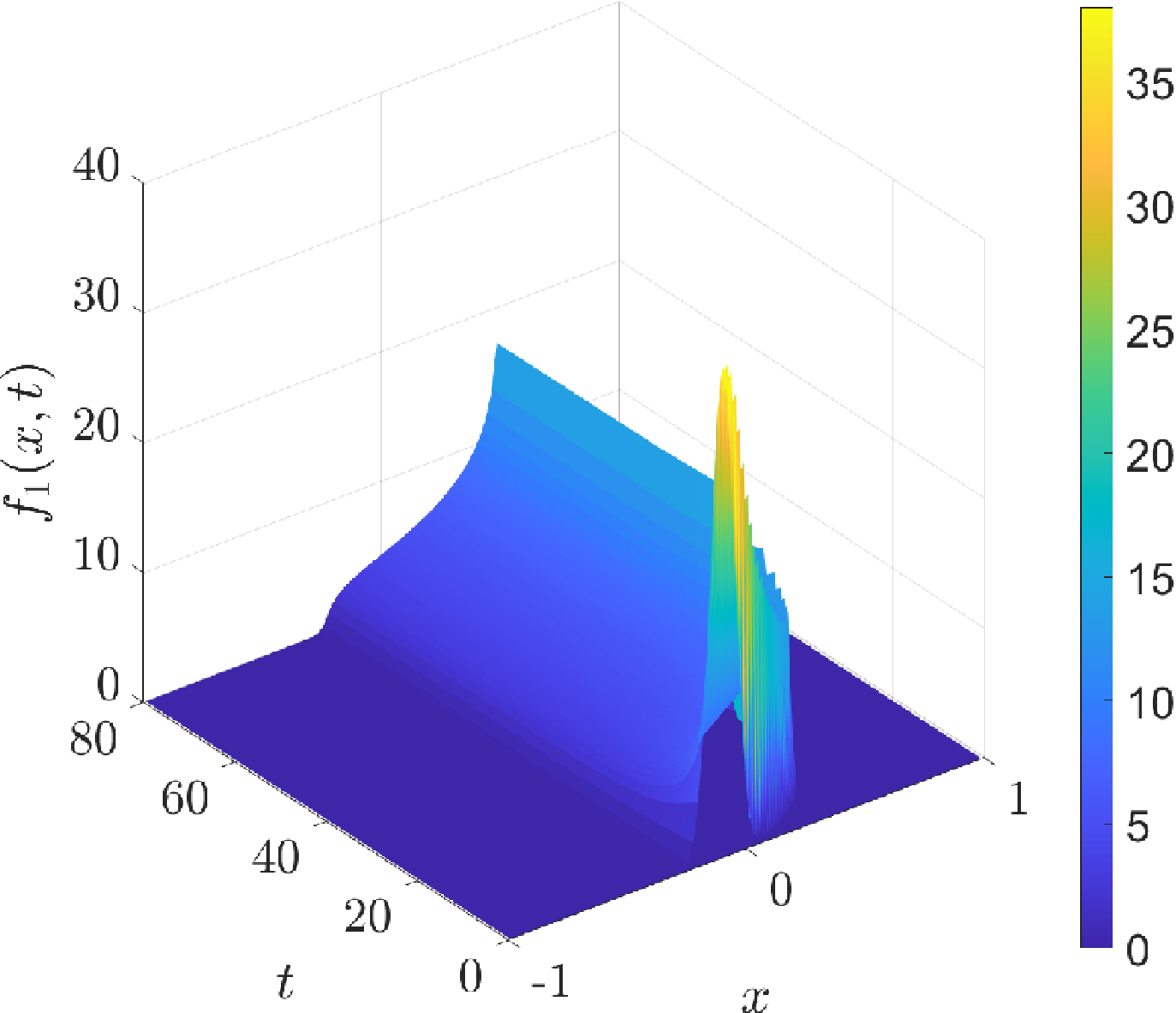}  						\includegraphics[width=0.4\textwidth]{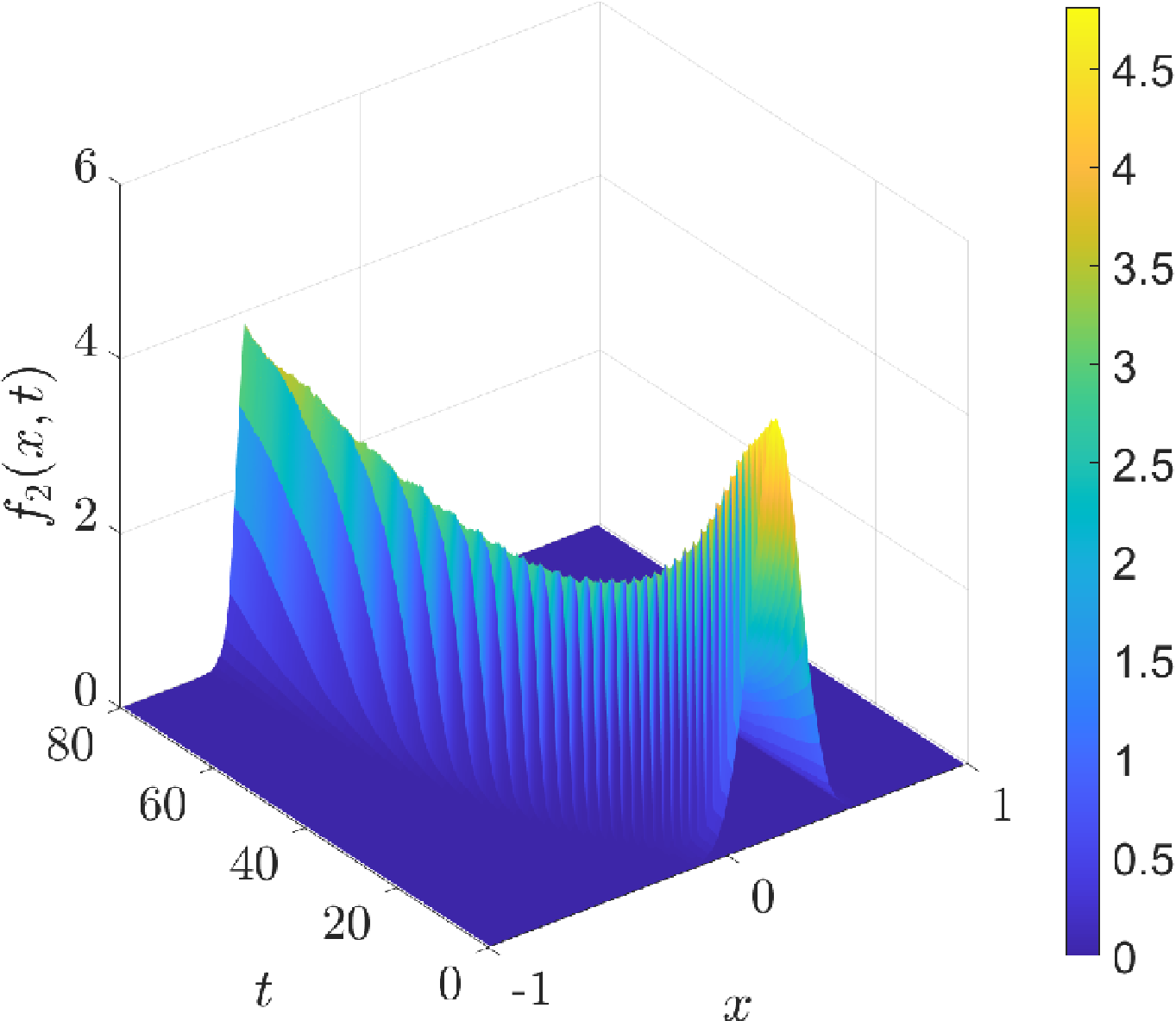}
	\caption{{Profiles of the opinions of the two populations at time $t=0$ (left) and at time $t=20$ (right).}}
	\label{fig:test_2pop3}
\end{figure*}

\begin{figure*}
	\centering
	\includegraphics[width=0.3\textwidth]{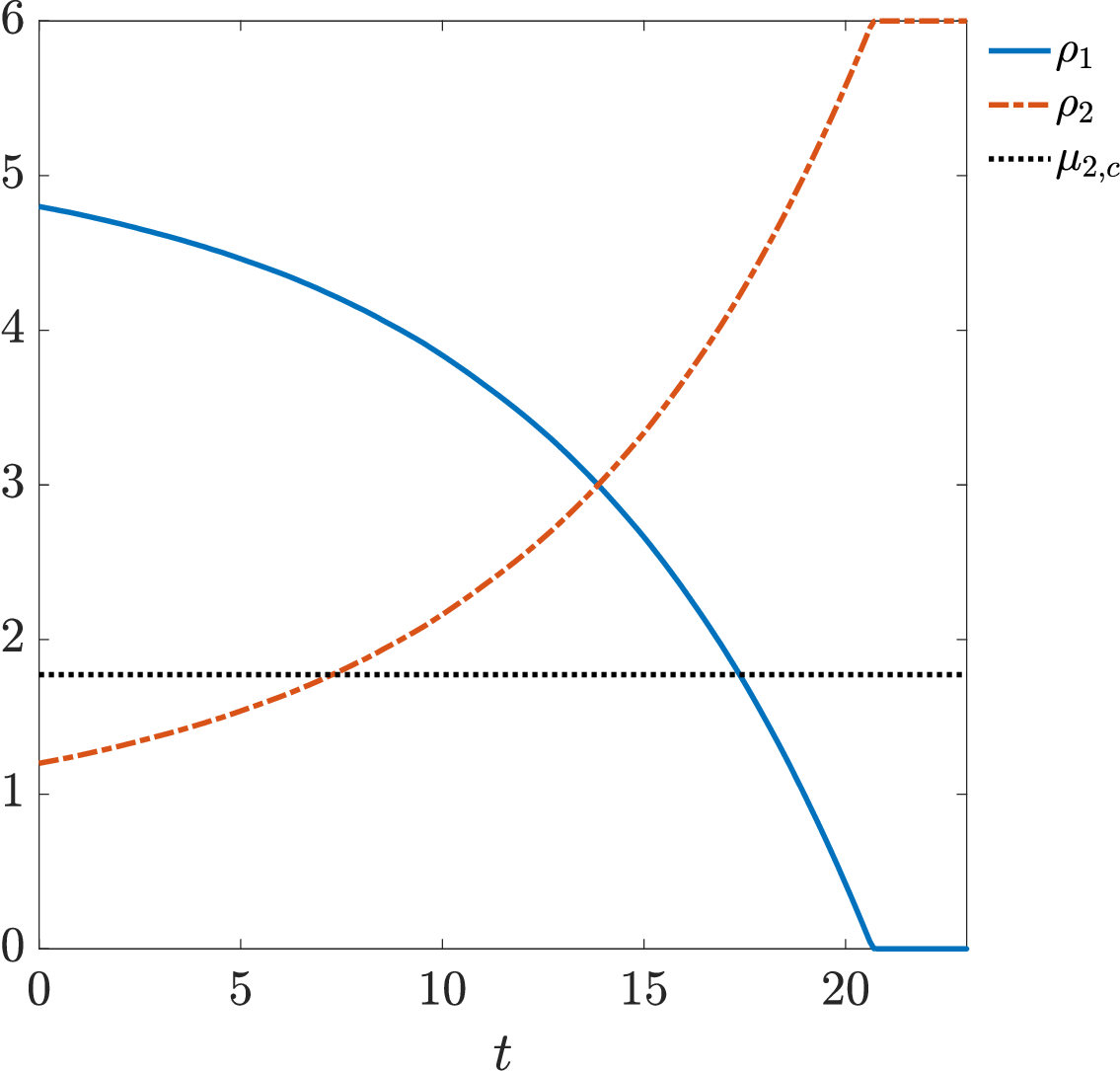}\quad
	\includegraphics[width=0.3\textwidth]{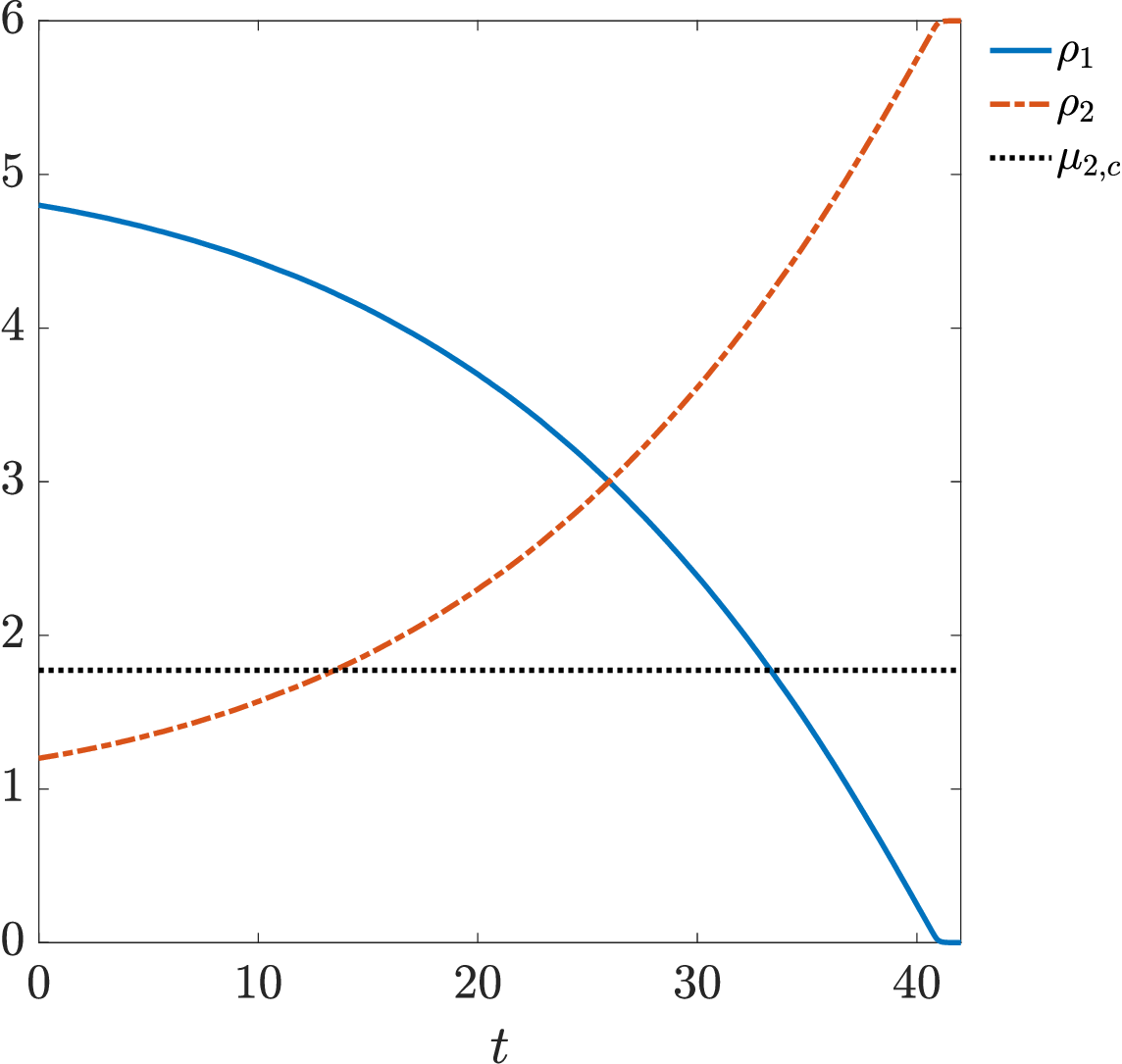}\quad \includegraphics[width=0.315\textwidth]{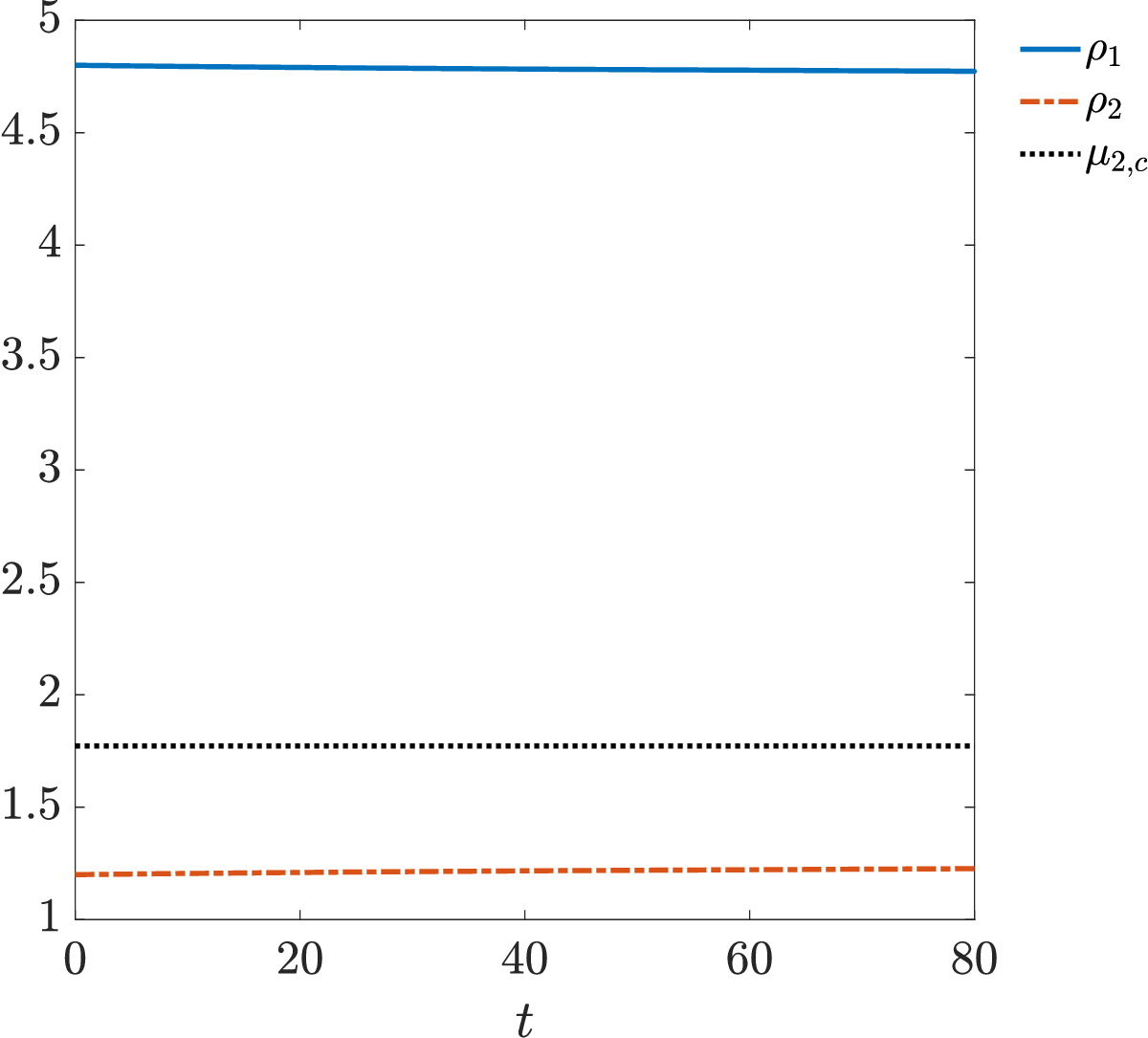}
	\caption{{Time evolution the masses $\rho_i$, $i=1,2$, of each population in time for the case  $\tau = 10^2$ (left), $\tau = 1$ (center), and for the case $\tau = 10^{-2}$ (right).}}
	\label{fig:test_2pop4}
\end{figure*}

In all the simulations, we use $N=10^6$ agents and $\epsilon = \min\{\tau/(1+\tau),0.1\}$. {Figure \ref{fig:test_2pop1} shows the results for $\tau=10^2$: on the left the distribution of opinions in time for the first group, on the right for the second group. In this setting, the evolution of opinion leading to a steady state is much slower than the migration of mass from one population to the other. The result is that many persons migrate from the first to the second group and then consensus around the mean value is created. Figure \ref{fig:test_2pop2} shows the evolution of $f_1(x,t)$ and $f_2(x,t)$ over time for $\tau=1$, meaning that the evolution of opinions and the migration of mass from one population to the other happens on the same time scale, while Figure \ref{fig:test_2pop3} shows the evolution of $f_1(x,t)$ and $f_2(x,t)$ for $\tau=10^{-2}$, meaning that the evolution of opinion is much faster than the migration of mass from one population to the other. Figure \ref{fig:test_2pop4}, finally, shows a comparison of the two masses $\rho_1$ and $\rho_2$ for the case  $\tau = 10^2$, $\tau = 1$, and $\tau = 10^{-2}$. We can see how for the case $\tau = 1$ the final shape is similar to the case with $\tau = 10^2$, even if the time evolution leads to different configurations.  For $\tau = 10^{-2}$ the result is different, even if the total mass of individuals belonging to one single group remains almost constant in time, there is some migration of individuals from the two groups and this has an impact over the final distribution that is spread over a larger support with respect to the previous case.}

\section*{Conclusion and perspectives}\label{sec:conc}
In this paper, we have explored a new path for modeling the impact of social groups on shaping opinions of individuals. The core idea on which the model is derived relies on the hypothesis that the size of the population plays a key role in the conformity process leading to strong polarization when a critical value is reached. The mathematical model which is obtained exhibits indeed a phase transition between a smooth and a singular solution at the equilibrium, sharing some common traits with the well-known Bose-Einstein condensation phenomenon. Specifically, we started from a microscopic interaction law between an individual and a background and then we upscaled our point of view through a mean field approximation and we finally obtained a Fokker-Planck-type of equation with quadratic drift  for the density of opinion.  We first explored the analytical properties of the model and then we studied its qualitative behavior through a series of numerical experiments with the scope of documenting its capability in describing certain well known patterns related to opinion formation, such as polarization and consensus effects. In the last part we briefly introduced a competing multi-population model with the scope of simulating the political competition among parties which is mainly based on this new conforming dynamics. In the future, we will concentrate on the possibility for the introduced modeling approach to describe blow-up in finite-time. Furthermore, we will   explore more in detail the multi-population model, focusing on the possibility to embed control strategies in the introduced equations.

\section*{Acknowledgments}
This work has been written within the activities of the GNCS and GNFM groups of INdAM. G.D. and M.Z.  acknowledge the support of the Italian Ministry of University and Research (MUR) through the PRIN 2020 project (No. 2020JLWP23) ``Integrated Mathematical Approaches to Socio–Epidemiological Dynamics". E.C. acknowledges the support of ``INdAM - GNCS Project", CUP E53C23001670001 and of MUR-PRIN Project 2022 PNRR (No. P2022JC95T) ``Data-driven discovery and control of multi-scale interacting artificial agent systems'', financed by the European Union - Next Generation EU. M.Z. acknowledges the support of the ICSC – Centro Nazionale di Ricerca in High Performance Computing, Big Data and Quantum Computing, funded by European Union - NextGeneration EU.

\end{document}